\documentclass[prd,preprint,tightenlines,showpacs,preprintnumbers,nofootinbib,eqsecnum,superscriptaddress]{revtex4-1}

 \usepackage[dvips,final]{graphicx}
  \usepackage{amssymb}
   \usepackage{amsmath}
    \usepackage{amsfonts}
     \usepackage{epsfig}
      \usepackage{bm}

\usepackage[section]{placeins}

\usepackage{sidecap}
\usepackage{multirow}
\usepackage{booktabs}
\usepackage{array}
\usepackage{tabularx}
\usepackage{xcolor}
\usepackage{pstricks}

\begin{document}

\newcommand{\ds}{\displaystyle}
\newcommand{\mc}{\multicolumn} 
\newcommand{\bce}{\begin{center}}
\newcommand{\ece}{\end{center}}
\newcommand{\beq}{\begin{equation}}
\newcommand{\eeq}{\end{equation}}
\newcommand{\bea}{\begin{eqnarray}}

\newcommand{\eea}{\end{eqnarray}}
\newcommand{\cont}{\nonumber\eea\bea}
\newcommand{\cl}[1]{\begin{center} {#1} \end{center}}
\newcommand{\ea}{\end{array}}

\newcommand{\ab}{{\alpha\beta}}
\newcommand{\cd}{{\gamma\delta}}
\newcommand{\dc}{{\delta\gamma}}
\newcommand{\ac}{{\alpha\gamma}}
\newcommand{\bd}{{\beta\delta}}
\newcommand{\abc}{{\alpha\beta\gamma}}
\newcommand{\eps}{{\epsilon}}
\newcommand{\lam}{{\lambda}}
\newcommand{\mn}{{\mu\nu}}
\newcommand{\mpnp}{{\mu'\nu'}}
\newcommand{\Amuu}{{A_{\mu}}}
\newcommand{\Amuo}{{A^{\mu}}}
\newcommand{\Vmuu}{{V_{\mu}}}
\newcommand{\Vmuo}{{V^{\mu}}}
\newcommand{\Anuu}{{A_{\nu}}}
\newcommand{\Anuo}{{A^{\nu}}}
\newcommand{\Vnuu}{{V_{\nu}}}
\newcommand{\Vnuo}{{V^{\nu}}}
\newcommand{\Fmnu}{{F_{\mu\nu}}}
\newcommand{\Fmno}{{F^{\mu\nu}}}

\newcommand{\abcd}{{\alpha\beta\gamma\delta}}


\newcommand{\bsigma}{\mbox{\boldmath $\sigma$}}
\newcommand{\beps}{\mbox{\boldmath $\varepsilon$}}
\newcommand{\btau}{\mbox{\boldmath $\tau$}}
\newcommand{\brho}{\mbox{\boldmath $\rho$}}
\newcommand{\bpipi}{\mbox{\boldmath $\pi\pi$}} 
\newcommand{\bss}{\bsigma\!\cdot\!\bsigma}
\newcommand{\btt}{\btau\!\cdot\!\btau}
\newcommand{\bnabla}{\mbox{\boldmath $\nabla$}}
\newcommand{\bphi}{\mbox{\boldmath $\tau$}}
\newcommand{\bvarphi}{\mbox{\boldmath $\rho$}}
\newcommand{\bE}{\mbox{\boldmath $E$}}
\newcommand{\bDelta}{\mbox{\boldmath $\Delta$}}
\newcommand{\bGamma}{\mbox{\boldmath $\Gamma$}}
\newcommand{\bpsi}{\mbox{\boldmath $\psi$}}
\newcommand{\bPsi}{\mbox{\boldmath $\Psi$}}
\newcommand{\bPhi}{\mbox{\boldmath $\Phi$}}
\newcommand{\bnab}{\mbox{\boldmath $\nabla$}}
\newcommand{\bpi}{\mbox{\boldmath $\pi$}}
\newcommand{\btheta}{\mbox{\boldmath $\theta$}}
\newcommand{\bkappa}{\mbox{\boldmath $\kappa$}}
\newcommand{\bgamma}{\mbox{\boldmath $\gamma$}}

\newcommand{\bp}{\mbox{\boldmath $p$}}
\newcommand{\ba}{\mbox{\boldmath $a$}}
\newcommand{\bq}{\mbox{\boldmath $q$}}
\newcommand{\br}{\mbox{\boldmath $r$}}
\newcommand{\bs}{\mbox{\boldmath $s$}}
\newcommand{\bk}{\mbox{\boldmath $k$}}
\newcommand{\bl}{\mbox{\boldmath $l$}}
\newcommand{\bb}{\mbox{\boldmath $b$}}
\newcommand{\be}{\mbox{\boldmath $e$}}
\newcommand{\bP}{\mbox{\boldmath $P$}}
\newcommand{\bV}{\mbox{\boldmath $V$}}
\newcommand{\bI}{\mbox{\boldmath $I$}}
\newcommand{\bJ}{\mbox{\boldmath $J$}}

\newcommand{\bT}{{\bf T}}
\newcommand{\fph}{${\cal F}$}
\newcommand{\aph}{${\cal A}$}
\newcommand{\dph}{${\cal D}$}
\newcommand{\fpi}{f_\pi}
\newcommand{\mpi}{m_\pi}
\newcommand{\Tr}{{\mbox{\rm Tr}}}
\def\Qb{\overline{Q}}
\newcommand{\delu}{\partial_{\mu}}
\newcommand{\delo}{\partial^{\mu}}
\newcommand{\up}{\!\uparrow}
\newcommand{\upup}{\uparrow\uparrow}
\newcommand{\updo}{\uparrow\downarrow}
\newcommand{\uu}{$\uparrow\uparrow$}
\newcommand{\ud}{$\uparrow\downarrow$}
\newcommand{\auu}{$a^{\uparrow\uparrow}$}
\newcommand{\aud}{$a^{\uparrow\downarrow}$}
\newcommand{\pu}{p\!\uparrow}
\newcommand{\qp}{quasiparticle}
\newcommand{\sa}{scattering amplitude}
\newcommand{\ph}{particle-hole}
\newcommand{\qcd}{{\it QCD}}
\newcommand{\integ}{\int\!d}
\newcommand{\ie}{{\sl i.e.~}}
\newcommand{\etal}{{\sl et al.~}}
\newcommand{\etc}{{\sl etc.~}}
\newcommand{\rhs}{{\sl rhs~}}
\newcommand{\lhs}{{\sl lhs~}}
\newcommand{\eg}{{\sl e.g.~}}
\newcommand{\ef}{\epsilon_F}
\newcommand{\sigt}{d^2\sigma/d\Omega dE}
\newcommand{\sige}{{d^2\sigma\over d\Omega dE}}
\newcommand{\rpaeq}{\beq
\left ( \begin{array}{cc}
A&B\\
-B^*&-A^*\end{array}\right )
\left ( \begin{array}{c}
X^{(\kappa})\\Y^{(\kappa)}\end{array}\right )=E_\kappa
\left ( \begin{array}{c}
X^{(\kappa})\\Y^{(\kappa)}\end{array}\right )
\eeq}

\newcommand{\ket}[1]{{#1} \rangle}
\newcommand{\bra}[1]{\langle {#1} }

\newcommand{\Bigket}[1]{{#1} \Big\rangle}
\newcommand{\Bigbra}[1]{\Big\langle {#1} }

\newcommand{\ave}[1]{\langle {#1} \rangle}
\newcommand{\Bigave}[1]{\left\langle {#1} \right\rangle}
\newcommand{\half}{{1\over 2}}

\newcommand{\singlespace}{
    \renewcommand{\baselinestretch}{1}\large\normalsize}
\newcommand{\doublespace}{
    \renewcommand{\baselinestretch}{1.6}\large\normalsize}
\newcommand{\bftau}{\mbox{\boldmath $\tau$}}
\newcommand{\bfalpha}{\mbox{\boldmath $\alpha$}}
\newcommand{\bfgamma}{\mbox{\boldmath $\gamma$}}
\newcommand{\bfxi}{\mbox{\boldmath $\xi$}}
\newcommand{\bfbeta}{\mbox{\boldmath $\beta$}}
\newcommand{\bfeta}{\mbox{\boldmath $\eta$}}
\newcommand{\bfpi}{\mbox{\boldmath $\pi$}}
\newcommand{\bfphi}{\mbox{\boldmath $\phi$}}
\newcommand{\bfR}{\mbox{\boldmath ${\cal R}$}}
\newcommand{\bfL}{\mbox{\boldmath ${\cal L}$}}
\newcommand{\bfM}{\mbox{\boldmath ${\cal M}$}}
\def\dblint{\mathop{\rlap{\hbox{$\displaystyle\!\int\!\!\!\!\!\int$}}
    \hbox{$\bigcirc$}}}
\def\ut#1{$\underline{\smash{\vphantom{y}\hbox{#1}}}$}

\def\UNITY{{\bf 1\! |}}
\def\Pom{{\bf I\!P}}
\def\lsim{\mathrel{\rlap{\lower4pt\hbox{\hskip1pt$\sim$}}
    \raise1pt\hbox{$<$}}}         
\def\gsim{\mathrel{\rlap{\lower4pt\hbox{\hskip1pt$\sim$}}
    \raise1pt\hbox{$>$}}}         

\newcommand\scalemath[2]{\scalebox{#1}{\mbox{\ensuremath{\displaystyle #2}}}}

\newcommand{\RP}[1]{{\blue RP: #1}}
\newcommand{\WS}[1]{{\red WS: #1}}

\title{Light-front approach to axial-vector quarkonium $\gamma^* \gamma^*$ form factors}

\author{Izabela Babiarz}
\email{izabela.babiarz@ifj.edu.pl}
\affiliation{Institute of Nuclear Physics, Polish Academy of Sciences, 
ul. Radzikowskiego 152, PL-31-342 Krak{\'o}w, Poland}

\author{Roman Pasechnik}
\email{roman.pasechnik@thep.lu.se}
\affiliation{Department of Astronomy and Theoretical Physics,
Lund University, SE-223 62 Lund, Sweden}

\author{Wolfgang Sch\"afer}%
\email{Wolfgang.Schafer@ifj.edu.pl}
\affiliation{Institute of Nuclear
Physics, Polish Academy of Sciences, ul. Radzikowskiego 152, PL-31-342 
Krak{\'o}w, Poland}

\author{Antoni Szczurek}
\email{antoni.szczurek@ifj.edu.pl}
\affiliation{Institute of Nuclear
Physics, Polish Academy of Sciences, ul. Radzikowskiego 152, PL-31-342 
Krak{\'o}w, Poland}
\affiliation{Faculty of Mathematics and Natural Sciences,
University of Rzesz\'ow, ul. Pigonia 1, PL-35-310 Rzesz\'ow, Poland\vspace{5mm}}

\begin{abstract}
\vspace{5mm}
In this work, we perform a detailed study of transition form factors for axial-vector meson production via the two-photon fusion process $\gamma^* \gamma^* \to 1^{++}$, with space-like virtual photons in the initial state and a $P$-wave axial-vector quarkonium in the final state. In this analysis, we employ the formalism of light-front quarkonium wave functions obtained from a solution of the Schr\"odinger equation for a selection of interquark potentials for $Q \bar Q$ interaction. We found the helicity structure and covariant decomposition of the matrix elements that can be generically applied for any $q \bar q$ axial-vector meson $\gamma^* \gamma^* \to 1^{++}$ transition, while our numerical results are given for the phenomenologically relevant charmonium $\chi_{c1}$ state. We present the helicity form factors as functions of both photon virtualities. We also obtain, that $Q F_{\rm LT}(Q^2,0)/F_{\rm TT}(Q^2,0) = {\rm const.}$
\end{abstract}

\maketitle

\section{Introduction}
\label{sec:intro}

Exclusive production in $e^+e^-$ collisions in $\gamma\gamma$ fusion has been considered as a probe for internal structure of QCD bounds states quantified by meson-photon transition form factors \cite{Chernyak:2014wra,Kopp:1973hp,Poppe:1986dq,Schuler:1997yw,Li:2021ejv}. Besides the QED production mechanism, such form factor can be used also in color-singlet QCD production mechanism via gluon-gluon fusion. The latter has significant implications in physics of vector and axial-vector meson production in hadronic and nuclear collisions, which recently attracts a lot of attention \cite{Brambilla:2010cs}. 

In Refs.~\cite{Babiarz:2019sfa,Babiarz:2019mag,Babiarz:2020jkh}, we have analysed production of scalar and pseudo-scalar mesons in the light-cone formulation which are $P$- and $S$-wave $Q\bar Q$ states, respectively. We present the results including the intrinsic (anti)quark motion in the meson wave function, and also taking the non-relativistic QCD (NRQCD) limit which in our case of the $P$-wave meson is proportional to the derivative of the wave function at the origin.

In this work, we continue our studies of the light-cone approach to the transition form factors focusing on $\gamma\gamma$ fusion into axial-vector $J^{PC}=1^{++}$ mesons. We utilise several well-known models for the interquark potentials in determination of the axial-vector meson wave function on the light cone. Our consideration is general for quarkonium states while our numerical analysis would be concentrated on $\chi_{c,J=1}$ state. The corresponding form factors are derived for various meson polarisation states. Provided that, due to the Landau-Yang theorem \cite{Landau:1948kw, Yang:1950rg}, the axial-vector meson does not decay into a pair of real photons, the corresponding transition form factors and hence the production helicity amplitudes become non-trivial only in the case of virtual initial-state photons.

Our formulas could easily be extended to light $f_1$ or $a_1$ mesons, adopting an appropriate light-front quark model. Here we restrict ourselves to heavy quarkonia in the numerical predictions as for light mesons the production is not fully under control \cite{Szczurek:2020hpc}.

The paper is organised as follows. In Sect.~\ref{sec:LC-formulation} we recall the basics of the LF approach to the axial-vector quarkonium and derive the 
$\gamma^* \gamma^* \to A$ amplitude in terms of the LFWF of the bound state.
In Sect.~\ref{sec:tensor-dec} we present a general covariant description of the $\gamma^* \gamma^* \to 1^{++}$ amplitude in terms of invariant helicity form factors which we relate to the transition form factors of our light cone amplitude. In Sect.~\ref{sec:ff_results} we show our numerical results for the relevant form factors for the case of the $\chi_{c1}$ meson. We will also dicuss the reduced width for the $\gamma^* \gamma$ decay and the $Q^2$ dependence of the $\gamma^* \gamma$ cross section relevant to single tagged $e^+ e^-$ collisions. Finally, concluding remarks and a summary are given in Sect.~\ref{sec:Conclusions}.

\section{Light-front formulation of the $\gamma^* \gamma^* \to 1^{++}$ process}
\label{sec:LC-formulation}

\subsection{Axial-vector meson wave function}

In order to construct the light-front wave function (LFWF) of the $Q \bar Q$ bound state of good angular momentum quantum numbers, we follow the procedure based on the Melosh transformation \cite{Melosh:1974cu,Jaus:1989au} and Terent'ev substitution \cite{Terentev:1976jk} starting from the quark-model rest-frame wave function (WF). In this work, we are particularly focused on the axial-vector $J^{PC} = 1^{++}$ quarkonium state (known as ${}^{2S+1}L_J = {}^3P_1$), whose WF depends on the meson spin projections $\lambda_A = \pm 1,0$, as well as on polarizations $\tau, \bar \tau$ of the heavy quark and antiquark (with mass $m_Q$):
\begin{eqnarray}
\Psi^{(\lambda_A)}_{\tau \bar \tau} (\vec k) &=&   \sum_{L_z+S_z = \lambda_A} Y_{1L_z}(\hat k)
\Bigbra{ \frac{1}{2} \frac{1}{2} \tau \bar \tau } | \Bigket{1 S_z}
\bra{11 L_z S_z} | \ket{1 \lambda_A} {u(k) \over k}
\nonumber \\
&=& {1 \over 2}  \sqrt{3 \over 4 \pi} \,  \xi^{\tau \dagger}_Q \, \, \Big(
\vec{\sigma} \cdot  {\vec{k} \times \vec{E}(\lambda_A) \over k} \Big )  i \sigma_2 \, \xi^{\bar \tau *}_{\bar Q} \,\,  {u(k) \over k} \, .
\label{eq:RF_WF}
\end{eqnarray}
Here, $\vec k$ is the three-momentum of the heavy quark in the rest frame of the $Q\bar Q$ pair, $k = |\vec k|$, $\vec E(\lambda_A)$ is the polarization vector of the axial-vector meson defined in Appendix~\ref{sec:states}, and $u(k)$ is the radial WF which we obtain from solutions of the Schr\"odinger equation for several different models of the quark-antiquark potential.

The Melosh transform of the spin-orbital part of the WF relies on the transition to LF spinors $\chi_Q^\lambda, \chi_{\bar Q}^{\bar \lambda *}$, by means of a rotation:
\begin{eqnarray}
 \xi_Q &=& R(z,\bk) \chi_Q \, , \qquad
 \xi^*_{\bar Q} = R^*(1-z,-\bk) \chi^*_{\bar Q} \, ,
\end{eqnarray}
with the unitary matrix $R(z,\bk)$ defined as
\begin{eqnarray}
R(z, \bk) = { m_Q + z M_{Q \bar Q} - i \vec \sigma \cdot ( \vec n \times \vec k) \over \sqrt{ (m_Q + z M_{Q \bar Q})^2 + \bk^2  }} = { m_Q + z M_{Q\bar Q} - i \vec \sigma \cdot ( \vec n \times \vec k) \over \sqrt{ zM_{Q \bar Q} (M_{Q \bar Q}+2m_Q) }} \, .
\end{eqnarray}
Here, $z$ and $1-z$ are the fractions of the meson's LF plus-momentum carried by the quark and antiquark, respectively, while their transverse momenta are denoted by $\pm \bk$. We also introduced the unit vector $\vec{n} = (0,0,1)$ such that the vector product reads, $\vec n \times \vec{k} = (-k_y,k_x,0)$. Furthermore, $M_{Q \bar Q}$ is the invariant mass of the $Q \bar Q$ system obtained from
\begin{eqnarray}
 M_{Q \bar Q}^2 = {\bk^2 + m_Q^2 \over z (1-z)} \,.
\end{eqnarray}
The LFWF then is obtained as follows
\begin{eqnarray}
 \Psi^{(\lambda_A)}_{\lambda \bar \lambda} (z,\bk) = \chi_Q^{\lambda \dagger}
 \, {\cal O}'_{\lambda_A} \, i \sigma_2 \, \chi_{\bar Q}^{\bar \lambda *}  \, \,
 \psi(z, \bk)  \, \sqrt{2 (M_{Q\bar Q}^2 - 4 m_Q^2)}\, ,
\label{eq:LFWF}
\end{eqnarray}
where we pull out a square-root factor to simplify formulas further on. 
The spin-orbital part is encoded in the $2 \times 2$-matrix,
\begin{eqnarray}
{\cal O}'_{\lambda_A} = \sqrt{3 \over 2} \,   R^\dagger(z,\bk) \Big( \vec \sigma \cdot  {\vec k \times \vec E(\lambda_A) \over \sqrt{2} k} \Big ) R(1-z,-\bk) \, .
\label{eq:Melosh}
\end{eqnarray}
The momentum three-vector in the rest frame of the pair is obtained in terms of LF variables as
\begin{eqnarray}
 \vec k = (\bk, k_z) = \Big(\bk, \half (2z-1) M_{Q \bar Q}\Big) \, .
\end{eqnarray}
Finally, the ``radial'' LFWF is related to its rest-frame counterpart as 
\begin{eqnarray}
\psi(z,\bk) = { \pi \sqrt{M_{Q \bar Q}} \over 2 \sqrt{2}} \, \,  {u(k) \over k^2} \, . 
\end{eqnarray}
Explicit expressions for the LFWF of Eq.~(\ref{eq:LFWF}) are found in Appendix~\ref{sec:states}.

\subsection{Master formula for the light-front helicity amplitude}

We now turn to a calculation of the $\gamma^* \gamma^*$-fusion amplitude. Analogously to our previous works \cite{Babiarz:2019mag,Babiarz:2019sfa,Babiarz:2020jkh}, we write down the amplitude in the frame where the photon momenta are $q_{1\mu} = q_{1+} n^+_\mu + q^\perp_{1\mu}$ and $q_{2\mu} = q_{2-} n^-_\mu  + q^\perp_{2\mu}$, with the light-like vectors $n^+_\mu, n^-_\mu$, which fulfill $n^+ \cdot n^- = 1$. It proves convenient to choose the off-shell polarization-vectors $n^+_\mu$ and $n^-_\mu$ for the first and second photon, respectively. Our amplitude can be obtained from the convolution of the $\gamma^* \gamma^* \to  Q \bar Q$ amplitude with the LFWF of the axial-vector meson with polarization $\lambda_A$:
\begin{eqnarray}
n^{+\mu} n^{-\nu} {\cal M}_{\mu \nu} (\gamma^* \gamma^* \to A(\lambda_A)) = {\cal N} \int \frac{dz d^2 \bk} {z(1-z)16 \pi^3}
\sum_{\lambda \bar \lambda} \Psi^{(\lambda_A)*}_{\lambda \bar \lambda}(z,\bk)  n^{+\mu} n^{-\nu} {\cal A}_{\mu \nu}^{\lambda \bar \lambda} \, ,
\end{eqnarray}
where we pulled out the fine structure constant $\alpha_{\rm em}$ and a normalization factor from the sum over colors, such that
\begin{eqnarray}
{\cal N} = 4 \pi \alpha_{\rm em} e_f^2 \, { \mathrm{Tr} \, \openone_{\mathrm{color}} \over \sqrt{N_c}} = 4 \pi \alpha_{\rm em} e_f^2 \sqrt{N_c} \, .
\end{eqnarray}
Here, $N_c = 3$, and the electric charge $e_f = 2/3$ for the charm quark of interest in this work. 

Our master formula \cite{Iza_thesis} for the light-front helicity amplitude of the $\gamma^* \gamma^* \to 1^{++}$ process reads
\begin{eqnarray}
&& {\cal T}(\lambda_A) \equiv \int \frac{dz d^2 \bk} {z(1-z)16 \pi^3}
\sum_{\lambda \bar \lambda} \Psi^{(\lambda_A)*}_{\lambda \bar \lambda} n^{+\mu} n^{-\nu} {\cal A}_{\mu \nu}^{\lambda \bar \lambda} = (-2) \int {dz\, d^2\bk \over \sqrt{z(1-z)} 16 \pi^3} \nonumber \\
&&\times \Big\{-m_Q \Big[ {1 \over \bl_A\,^2 + \varepsilon^2} 
- {1 \over \bl_B\,^2 + \varepsilon^2}\Big] 
\Big( \sqrt{2}(\be(-)\bq_1) \Psi^{(\lambda_A) *}_{++}(z,\bk) 
+ \sqrt{2} (\be(+)\bq_1) \Psi^{(\lambda_A)*}_{--}(z,\bk) \Big) \nonumber\\
&&+ \Big( 2 z(1-z) Q_1^2 + (1-2z) (\bk \cdot \bq_1) \Big) 
\Big[ {1 \over \bl_A\,^2 + \varepsilon^2} - {1 \over \bl_B\,^2 + \varepsilon^2}\Big]
\Big( \Psi^{(\lambda_A)*}_{+-}(z,\bk) 
+ \Psi^{(\lambda_A)*}_{-+}(z,\bk) \Big) \nonumber \\
&&- (1-2z) (\bq_1 \cdot \bq_2) \Big[ {1-z \over \bl_A\,^2 + \varepsilon^2} + {z \over \bl_B\,^2 + \varepsilon^2} \Big] \Big( \Psi^{(\lambda_A)*}_{+-}(z,\bk) 
+ \Psi^{(\lambda_A)*}_{-+}(z,\bk) \Big) \nonumber\\
&&+ i [\bk,\bq_1]  \Big[ {1 \over \bl_A\,^2 + \varepsilon^2} - {1 \over \bl_B\,^2 + \varepsilon^2}\Big]\Big( \Psi^{(\lambda_A) *}_{+-}(z,\bk) 
- \Psi^{(\lambda_A) * }_{-+}(z,\bk) \Big) \nonumber \\
&&+ i [\bq_1,\bq_2]  \Big[ {1-z \over \bl_A\,^2 + \varepsilon^2} + {z \over \bl_B\,^2 + \varepsilon^2} \Big] \Big( \Psi^{(\lambda_A)* }_{+-}(z,\bk) 
- \Psi^{(\lambda_A)*}_{-+}(z,\bk) \Big) 
\Big\}\, .
\label{eq:general_heli}
\end{eqnarray}
Here, we have introduced the shorthand notation,
\begin{eqnarray}
\bl_A = - \bk + (1-z) \bq_2 \, , \qquad 
\bl_B = \bk + z \bq_2 \,,  \qquad Q_1^2 \equiv \bq_1^2 \,,
\end{eqnarray}
and 
\begin{eqnarray}
\varepsilon^2 = m_Q^2 + z(1-z) Q_1^2 \,.
\end{eqnarray}
We now wish to derive a representation of the helicity amplitudes in terms of the radial LFWFs, which can be readily used in a numerical analysis. To this end, we introduce an abbreviated notation for integrals over the radial LFWF:
\begin{eqnarray}
\ave{f(z,\bk)} \equiv (-2 ) {\sqrt {3\over 2}} 
\int {dz d^2\bk \over z(1-z) 16 \pi^3}  \psi(z,\bk)  \, f(z,\bk) \, ,
\end{eqnarray}
as well as
\begin{eqnarray}
{\cal A} = {1 \over \bl_A^2 + \varepsilon^2} - {1 \over \bl_B^2 + \varepsilon^2} \, , \qquad 
{\cal B} = {1-z \over \bl_A^2 + \varepsilon^2} + {z \over \bl_B^2 +\varepsilon^2 } \, . 
\end{eqnarray}

\subsection{Longitudinally polarised meson: $\lambda_A = 0$}

Starting from the master formula (\ref{eq:general_heli}), one straightforwardly obtains the helicity amplitude for $\lambda_A = 0$ state production:
\begin{eqnarray}
{\cal T}(0) &=& (-2i) \sqrt{{3 \over 2}} \int {dz d^2\bk \over z(1-z) 16 \pi^3}  \psi(z,\bk) \Big \{ 
{4 m_Q^2 \over M_{Q \bar Q}} [\bq_1,\bk] {\cal A} + {4 \bk^2 \over M_{Q \bar Q}} [\bq_1,\bk] {\cal A} \nonumber \\
&+& [\bq_1,\bq_2] {- 4 \bk^2 \over M_{Q \bar Q}} {\cal B}  \Big\} = i [\bq_1,\bq_2] I_0 + i [\bq_1,\bI_1] \, ,
\end{eqnarray}
where
\begin{eqnarray}
I_0 = -4 \,  \Bigave{{\bk^2 \over M_{Q \bar Q}} \, {\cal B}} \, , \qquad 
\bI_1 = 4 \Bigave{ \bk \, {m_Q^2 + \bk^2 \over M_{Q \bar Q}} \, {\cal A}} \,.
\end{eqnarray}
Now, by rotational invariance, we can write
\begin{eqnarray}
\bI_1 = \bq_2 \, I_1 \, , \qquad
I_1 = {\bq_2 \cdot \bI_1 \over Q_2^2} = {4 \over Q_2^2} \Bigave{ (\bq_2 \cdot \bk) \, {m_Q^2 + \bk^2 \over M_{Q\bar Q} }
\, {\cal A}} \, , \qquad Q_2^2 \equiv  \bq_2^2 \,.
\end{eqnarray}
We remind the reader that $M_{Q \bar Q}$ depends both on $z$ and $\bk^2$.
Finally, our amplitude reads
\begin{eqnarray}
{\cal T}(0) = i [\bq_1,\bq_2] \Big( I_0(Q_1^2, Q_2^2) + I_1 (Q_1^2,Q_2^2) \Big ) \equiv 
i [\bq_1,\bq_2] \,  \Phi_0(Q_1^2,Q_2^2) \,. 
\label{eq:def_Phi_0}
\end{eqnarray}
We can write out the explicit expressions for integrals $I_0,I_1$:
\begin{eqnarray}
I_0(Q_1^2,Q_2^2) &=& 8 \sqrt{3 \over 2} \int {dz d^2\bk \over z (1-z) 16 \pi^3} \, \psi(z,\bk) \, {\bk^2 \over M_{Q \bar Q}} \Big( {1 -z \over \bl_A^2 + \varepsilon^2} + { z \over \bl_B^2 + \varepsilon^2 } \Big ) \, , 
\nonumber \\
I_1(Q_1^2,Q_2^2) &=& - 8 \sqrt{3 \over 2} {1 \over Q_2^2} \int {dz d^2\bk \over z(1-z) 16 \pi^3} \psi(z,\bk) {\bk^2 + m_Q^2 \over M_{Q\bar Q} } \, (\bq_2 \cdot \bk) \Big( 
{ 1 \over \bl_A^2 + \varepsilon^2} - {1 \over \bl_B^2 + \varepsilon^2}
\Big) \, . \nonumber \\
\label{eq:Integrals_Phi_0}
\end{eqnarray}

\subsection{Transversely polarised meson: $\lambda_A = \pm 1$} 

In the case of transversely polarised axial-vector meson, the master formula (\ref{eq:general_heli}) leads to
\begin{eqnarray}
{\cal T}(\pm) &=& (-4 i) \sqrt{3 \over 2} \int {dz d^2\bk \over z(1-z) 16 \pi^3}  \psi(z,\bk)
\Big\{ [\bq_1,\bE^*] m_Q^2 (1-2z) {\cal A} +  2 Q_1^2 z(1-z) [\bk,\bE^*] {\cal A}  \nonumber \\
&-& (1-2z) (\bq_1 \cdot \bq_2) [\bk, \bE^*] {\cal B}
-(1-2z) [\bq_1,\bq_2]  (\bk \cdot \bE^*) {\cal B} 
+ (1-2z) (\bk \cdot \bq_1)  [\bk,\bE^*] {\cal A} \nonumber \\
&-& (1-2z) [\bk,\bq_1] (\bk \cdot \bE^*)  {\cal A} \Big\} \,.
\end{eqnarray}
Here, we have integrals that involve the vector $\bk$ as well as the tensor $k_i k_j$.

Let us introduce the following integrals,
\begin{eqnarray}
J_0 &=& 4m_Q^2 \ave{(1-2z) \,  {\cal A}} \, , \qquad
{\bJ_1} = 8 \ave{z (1-z) \, \bk \, {\cal A}}\, , \nonumber \\
\bJ_2 &=& 4 \ave{(1-2z) \, \bk  \, {\cal B} }\, , \qquad
J_{ij} = 4 \ave{(1-2z) k_i k_j \, {\cal A} }\, .
\end{eqnarray}
Then, our amplitude takes the form,
\begin{eqnarray}
{\cal T}(\pm) &=& [\bq_1,\bE^*] \, J_0 + \bq_1^2 [\bJ_1,\bE^*] - (\bq_1 \cdot \bq_2) [\bJ_2,\bE^*]
- [\bq_1,\bq_2] (\bJ_2 \cdot \bE^*) \nonumber \\
&+& J_{ij} \varepsilon_{jl} \Big( q_{1i} E^*_l - E^*_i q_{1l} \Big) \, . \label{Tpm}
\end{eqnarray}
Again, we can simplify this structure by introducing scalar integrals:
\begin{eqnarray}
J_1 &=& {\bq_2 \cdot \bJ_1 \over Q_2^2} = {8 \over Q_2^2} \,  \ave{z (1-z) \, (\bq_2 \cdot \bk) \, {\cal A}} \, , \nonumber \\
J_2 &=&  {\bq_2 \cdot \bJ_2 \over Q_2^2} = {4 \over Q_2^2} \, \ave{(1-2z) \, (\bq_2 \cdot \bk)  \, {\cal B} } \, . \label{J-12}
\end{eqnarray}
The tensor integral can be decomposed into two scalar functions,
\begin{eqnarray}
J_{ij} = \Big( \delta_{ij} - {q_{2i}q_{2j} \over Q_2^2} \Big) \, J_3 + {q_{2i}q_{2j} \over Q_2^2} \, J_4 \, ,
\end{eqnarray}
where 
\begin{eqnarray}
J_3 = {4 \over Q_2^2} \, \Bigave{(1-2z) \Big(Q_2^2 \bk^2 - ( \bq_2 \cdot \bk)^2 \Big) \, {\cal A} } \, , \quad
J_4 = {4 \over Q_2^2} \, \Bigave{(1-2z) ( \bq_2 \cdot \bk)^2  \, {\cal A} } \, . \label{J-34}
\end{eqnarray}
Then our amplitude becomes
\begin{eqnarray}
{\cal T}(\pm) &=& [\bq_1,\bE^*] (J_0 + 2 J_3) 
+ [\bq_2,\bE^*] Q_1^2 J_1 \nonumber \\
&+& \Big( (\bq_1 \cdot \bq_2) [\bq_2,\bE^*] 
+ [\bq_1,\bq_2] (\bq_2,\bE^*) \Big)  {1 \over Q_2^2}  \Big((J_4 - J_3) - Q_2^2 J_2 \Big) \, .
\end{eqnarray}
It can be further simplified by noticing that
\begin{eqnarray}
q_{1i} q_{2j} q_{2k} E^*_l \Big( \delta_{ij} \varepsilon_{kl} + \delta_{jl} \varepsilon_{ik} \Big) = q_{1i} q_{2j} q_{2k} E^*_l \Big( \delta_{ij} \varepsilon_{kl} - \delta_{lj} \varepsilon_{ki} \Big)\, .
\end{eqnarray}
The tensor in brackets is antisymmetric in $il$ and only the symmetric part in $kj$ enters. But in two dimensions, any antisymmetric in $il$ tensor must be proportional to $\varepsilon_{il}$. One can easily check that in the above expression one can replace
\begin{eqnarray}
 \delta_{ij} \varepsilon_{kl} - \delta_{lj} \varepsilon_{ki}  \rightarrow \,\varepsilon_{il} \delta_{jk} \,,
\end{eqnarray}
thus, yielding a useful identity
\begin{eqnarray}
(\bq_1 \cdot \bq_2) [\bq_2,\bE^*] 
+ [\bq_1,\bq_2] (\bq_2,\bE^*)  = [\bq_1,\bE^*] Q_2^2 \, .
\label{eq:vector_identity}
\end{eqnarray}
The latter enables us to represent our amplitude in the following compact form,
\begin{eqnarray}
{\cal T}(\pm) = [\bq_1,\bE^*] \, \Phi_1(Q_1^2,Q_2^2) + [\bq_2, \bE^*]  \, \Phi_2(Q_1^2,Q_2^2)  \, ,
\end{eqnarray}
with two form factors
\begin{eqnarray}
\Phi_1 = J_0 + J_3 + J_4 - Q_2^2 J_2  \, , \qquad 
\Phi_2 = Q_1^2 \, J_1 \, ,
\end{eqnarray}
which, using the expressions in Appendix \ref{sec:integrals} can be represented in the following integral form
\begin{eqnarray}
\Phi_1(Q_1^2,Q_2^2) &=& -4 \sqrt{3 \over 2} \int {dz d^2 \bk \over z (1-z) 16 \pi^3} \psi(z,\bk) (1-2z) \Big\{ (\bk^2 + m_Q^2) \Big( {1 \over \bl_A^2 + \varepsilon^2} - { 1 \over \bl_B^2 + \varepsilon^2} \Big) \nonumber \\
&&- (\bq_2 \cdot \bk) \Big(  {1-z \over \bl_A^2 + \varepsilon^2} + { z \over \bl_B^2 + \varepsilon^2}
\Big) \Big \} \, , \\
\Phi_2(Q_1^2,Q_2^2) &=& -8 \sqrt{3 \over 2} {Q_1^2 \over Q_2^2} \int {dz d^2 \bk \over z (1-z) 16 \pi^3}  \psi(z,\bk) z(1-z) (\bq_2 \cdot \bk) 
\Big( {1 \over \bl_A^2 + \varepsilon^2} - { 1 \over \bl_B^2 + \varepsilon^2} \Big)\,. \nonumber
\label{eq:Phi_1_Phi_2}
\end{eqnarray}
We thus have completed our first task and expressed our amplitude for the $\gamma^* \gamma^* \to A(\lambda_A)$ process in terms of form factors that are calculated as integrals involving the LFWF.

\subsection{NRQCD limit}

In the nonrelativistic QCD (NRQCD) limit, we expand the integrand of our amplitude to linear order in $\xi = z- {1 \over 2}$ and $\bk$. Then,
\begin{eqnarray}
z = \half + \xi \, , \quad 1-z = \half - \xi \, , \quad 
z(1-z) = {1 \over 4} - \xi^2 \sim {1 \over 4} \, .
\end{eqnarray}
We introduce $ \mu^2 \equiv (Q_1^2 + Q_2^2 + 4 m_Q^2)/4$, and expand the  relevant combinations of denominators:
\begin{eqnarray}
{\cal A} &=& {1 \over \bl^2_A + \varepsilon^2} - {1 \over \bl^2_B + \varepsilon^2} = {2 \bq_2 \cdot(\bk + \xi \bq_2) \over \mu^4} \,, \qquad
{\cal B} = {1-z \over \bl^2_A+ \varepsilon^2} + {z \over \bl_B^2 + \varepsilon^2} = {1 \over \mu^2} \, . \nonumber \\
\end{eqnarray}
Inserting these expressions into our master formula, we obtain the amplitude in the following useful form,
\begin{eqnarray}
{\cal T} &=& {-2 \over \mu^4} \int {dz d^2 \bk \over \sqrt{z (1-z)} 16 \pi^3} 
\Big\{ i [\bq_1,\bq_2] \mu^2 \,  \Big (\Psi^{(\lambda_A)*}_{+-}(z,\bk) -
\Psi^{(\lambda_A)*}_{-+}(z,\bk)
\Big ) \nonumber \\
&+& \Big( Q_1^2 (\bq_2 \cdot \bk) + \xi Q_1^2 Q_2^2 - 2 \xi \mu^2 (\bq_1\cdot \bq_2) \Big)
\Big (\Psi^{(\lambda_A)*}_{+-}(z,\bk) +
\Psi^{(\lambda_A)*}_{-+}(z,\bk)
\Big ) \\
&-& 2m_Q (\bq_2 \cdot \bk + \xi \bk^2) 
\Big( \sqrt{2} (\be^*(-)\cdot \bq_1) \Psi^{(\lambda_A)*}_{++}(z,\bk) + \sqrt{2} (\be^*(+)\cdot \bq_1)
\Psi^{(\lambda_A)*}_{--}(z,\bk)
 \Big) \Big \} \, . \nonumber
\end{eqnarray}
After somewhat lengthy calculations, we can reduce our result to a simple form:
\begin{eqnarray}
{\cal T}(0) &=& i [\bq_1,\bq_2] \, t_0 \,  {Q_1^2 + Q_2^2 \over [ Q_1^2 + Q_2^2 + 4m_Q^2]^2 } , \nonumber \\
{\cal T}(\pm) &=& i \Big( Q_1^2 [\be^*, \bq_2] - Q_2^2 [\be^*,\bq_1] \Big) \, t_0 \, { 2m_Q \over [ Q_1^2 + Q_2^2 + 4m_Q^2]^2 } \, , 
\end{eqnarray}
with
\begin{eqnarray}
t_0 \equiv 16 \sqrt{3 \over 2} \, {R'(0) \over \sqrt{\pi M^3}} \, = 4 \, R'(0) \sqrt{3 \over \pi m_Q^3}  \, ,
\end{eqnarray}
where the mass of the resonance $M = 2m_Q$ in the NRQCD approximation. 
We can now easily read off the explicit expressions for the form factors $\Phi_{0,1,2}(Q_1^2,Q_2^2)$ in the NRQCD limit:
\begin{eqnarray}
\Phi_0(Q_1^2, Q_2^2) &=& t_0 { Q_1^2 + Q_2^2 \over [Q_1^2 + Q_2^2 + 4m_Q^2]^2 }\, ,
\nonumber \\
\Phi_1(Q_1^2,Q_2^2) &=& t_0 \, {Q_2^2  \, 2 m_Q \over[Q_1^2 + Q_2^2 + 4m_Q^2]^2 }\, , 
\nonumber \\
\Phi_2(Q_1^2,Q_2^2) &=& t_0 \, {-Q_1^2 \, 2m_Q \over[Q_1^2 + Q_2^2 + 4m_Q^2]^2 } \, .
\label{eq:Phi_NRQCD}
\end{eqnarray}

\section{Tensor decomposition of the $\gamma^* \gamma^* \to 1^{++}$ amplitude}
\label{sec:tensor-dec}

\subsection{Helicity form factors}

Let us start by parametrizing the covariant $\gamma^* \gamma^* \to 1^{++}$ amplitude by three separately gauge invariant tensor structures with the corresponding invariant form factors $F_{\rm TT}$, $F_{\rm LT}$, and $F_{\rm TL}$: 
\begin{eqnarray}
{1 \over 4 \pi \alpha_{\rm em}}{\cal M}_{\mu \nu \rho} &=& i \Big( q_1 - q_2 + {Q_1^2 - Q_2^2 \over (q_1 + q_2)^2} (q_1 + q_2) \Big)_\rho  \, \tilde G_{\mu \nu} \, { M \over 2 X} F_{\rm TT}(Q_1^2,Q_2^2) \nonumber\\
&+& i e_\mu^L(q_1) \tilde G_{\nu \rho}   \, { 1\over \sqrt{X}} F_{\rm LT}(Q_1^2,Q_2^2)
+ i e_\nu^L(q_2) \tilde G_{\mu \rho}  \, { 1 \over \sqrt{X}} F_{\rm TL}(Q_1^2,Q_2^2) \, .
\label{eq:formfactors}
\end{eqnarray}
Above we introduced 
\begin{eqnarray}
\tilde G_{\mu \nu} = \varepsilon_{\mu \nu \alpha \beta} q_1^\alpha q_2^\beta \, ,
\end{eqnarray}
and the polarization vectors of longitudinal photons
\begin{eqnarray}
e_\mu^L(q_1) &=& \sqrt{ {-q_1^2 \over X}} \Big (q_{2\mu} - { q_1 \cdot q_2 \over q_1^2} q_{1\mu} \Big)\, , \qquad
e_\nu^L(q_2) = \sqrt{ {-q_2^2 \over X}} \Big (q_{1\nu} - { q_1 \cdot q_2 \over q_2^2} q_{2\nu} \Big)\, .
\end{eqnarray}
We are interested in the spacelike virtualities of photons only, i.e. $Q_i^2 = -q_i^2 = \bq_i^2 > 0, \, i = 1,2$, and we denote the standard kinematic variable
\begin{eqnarray}
X = (q_1\cdot q_2)^2 - q_1^2 q_2^2
= {1 \over 4} \Big( M^4 + 2 M^2(Q_1^2+Q_2^2) + (Q_1^2 - Q_2^2)^2 \Big) > 0 \, .
\end{eqnarray}

In Eq.~(\ref{eq:formfactors}) we have introduced kinematic factors in such a way, that the form factors $F_{\rm TT}, F_{\rm LT}, F_{\rm TL}$ all have the same mass-dimension as the invariant amplitude and are related to helicity amplitudes in the $\gamma^* \gamma^*$-c.m. frame in a straightforward manner without additional kinematic factors, for a more detailed discussion, see Appendix \ref{sec:helicity_cms}. Different conventions and tensor structures can be useful depending on the application, see e.g.~Refs.~\cite{Poppe:1986dq,Pascalutsa:2012pr,Kopp:1973hp,Milstein:2019yvz,Hoferichter:2020lap}.
Then, the amplitude for the final-state meson with polarization $\lambda_A$ is obtained as
\begin{eqnarray}
{\cal M}_{\mu \nu}(\lambda_A) = {\cal M}_{\mu \nu \rho} E^{\rho *}(\lambda_A) \, ,
\end{eqnarray}
in terms of the meson polarization four-vector $E^{\rho}(\lambda_A)$. All three tensor structures in (\ref{eq:formfactors}) are separately orthogonal to the four-momentum $P_\rho = (q_1 + q_2)_\rho$ of the meson. 
Here we are interested only in the case of the on-shell meson, $P^2 = M^2$. For the case of $P^2 \neq M^2$, the form factors must be regarded as a function of all virtualities, $Q_1^2,Q_2^2,P^2$ and in addition an independent tensor structure $\propto P_\rho \tilde G_{\mu \nu}$ appears.
The associated form factor is the ``anomaly form factor'' of the vector-vector-axial correlator. As we have $P\cdot E^*(\lambda_A) = 0$ for the physical state, it does not contribute to our problem.

In what follows, we choose a basis of polarization states for the axial-vector which is convenient for our light-front calculation of the relevant amplitudes. Such polarization states read explicitly:
\begin{eqnarray}
E_\rho(0) &=& {1 \over M} P_\rho - {M \over P_+} n^-_\rho \, , \qquad
E_\rho(\pm) = E^\perp_\rho(\pm) - {E^\perp(\pm) \cdot P \over P_+} n^-_\rho \, .
\end{eqnarray}
Here, the vector $E^\perp_\mu$ has only transverse components\footnote{We use the light-front parametrization of four-momenta $a= (a_+,a_-,\ba), a^2 = 2 a_+ a_- - \ba^2$.} :
\begin{eqnarray}
E^\perp(\pm) = ( 0,0, \bE(\pm)) \quad {\rm with} \quad 
\bE(\lambda) = - {1 \over \sqrt{2}} ( \lambda \be_x + i \be_y) \,.
\end{eqnarray}

Symmetry under exchanging $q_1 \leftrightarrow q_2 \, , \, \mu \leftrightarrow \nu$ (crossing symmetry) entails that our form factors must satisfy
\begin{eqnarray}
F_{\rm TT} (Q_1^2, Q_2^2) = - F_{\rm TT}(Q_2^2, Q_1^2)\, , \qquad 
F_{\rm LT}(Q_1^2,Q_2^2) =- F_{\rm TL}(Q_2^2, Q_1^2) \,.
\end{eqnarray}
Due to its antisymmetry, the form factor $F_{\rm TT}$ vanishes on the diagonal $F_{\rm TT} (Q^2,Q^2) =0$, and in particular for two on-shell photons $F_{\rm TT}(0,0)=0$, thereby fulfilling the Landau--Yang theorem.
The latter has no bearing on the behaviour of $F_{\rm LT}, F_{\rm TL}$. However, when the virtuality of the longitudinal photon goes to zero, we have
\begin{eqnarray}
F_{\rm LT} (Q_1^2,Q_2^2) \propto Q_1 \, , \qquad {\rm for} \, \, Q_1^2 \to 0 \, , 
\end{eqnarray}
and accordingly for $F_{\rm TL}$, in order to cancel a spurious singular behaviour of the tensor decomposition of ${\cal M}_{\mu \nu \rho}$. Note, that the limit
\begin{eqnarray}
f_{\rm LT} = \lim_{Q^2 \to 0} {F_{\rm LT} (Q^2,0) \over Q} \, ,
\end{eqnarray}
is therefore finite and generally nonzero. In absence of a finite decay width to $\gamma \gamma$ it serves as a measure of the two-photon coupling strength of the axial-vector meson.

\subsection{Projection on the light-front directions}

We now want to calculate the projections of the amplitude on the light-front directions $n^{+\mu} n^{-\nu}$. We are interested in the frame, where photon momenta have the form
\begin{eqnarray}
q_{1\mu} = q_1^+ n^+_\mu + q^\perp_{1\mu} \,, \qquad q_{2\mu} = q_2^- n^-_\mu + q^\perp_{2\mu} \,,
\end{eqnarray} 
such that $Q_i^2 = - q_i^{\perp2} = \bq_i^2$. Let us collect a few useful contractions:
\begin{eqnarray}
n^{+\mu} n^{-\nu} \tilde G_{\mu \nu} &=& \varepsilon^\perp_{\alpha \beta} q_1^\alpha q_2^\beta = 
[\bq_1, \bq_2] \, , \quad  n^+ \cdot e^L(q_1) = {Q_1 \over \sqrt{X}} q_2^- \, , \quad
n^- \cdot  e^L(q_2) = {Q_2 \over \sqrt{X}} q_1^+ \, .
\end{eqnarray}
Our amplitude now takes the form
\begin{eqnarray}
{1 \over 4 \pi \alpha_{\rm em}} n^{+\mu} n^{-\nu} {\cal M}_{\mu \nu \rho} &=& i [\bq_1,\bq_2] 
\Big(   q_1 - q_2 + {Q_1^2 - Q_2^2 \over M^2} (q_1 + q_2)
\Big)_\rho  \,  {M \over 2 X} F_{\rm TT}(Q_1^2, Q_2^2)
\nonumber \\
&+& i {q_1^+ q_2^- \over X} \Big( \varepsilon^\perp_{\rho \alpha} q^{\perp \alpha}_{2} Q_1 F_{\rm LT}(Q_1^2,Q_2^2) + 
\varepsilon^\perp_{\rho \alpha} q^{\perp \alpha}_{1} Q_2 F_{\rm TL}(Q_1^2,Q_2^2) \Big) \nonumber \\
&+& i {[\bq_1,\bq_2] \over X} \Big( Q_2 q_1^+ n^+_\rho F_{\rm TL}(Q_1^2,Q_2^2) - Q_1 q_2^- n_\rho^- F_{\rm LT}(Q_1^2,Q_2^2) \Big) \, . \end{eqnarray}
Furthermore, we obtain for the helicity amplitudes 
\begin{eqnarray}
{1 \over 4 \pi \alpha_{\rm em}} {\cal M}(0) &=& n^{+\mu} n^{-\nu} {\cal M}_{\mu \nu \rho} E^{\rho *}(0) \nonumber \\
&=& -i [\bq_1,\bq_2] \Big\{ \Big( M^2 + Q_1^2 - Q_2^2  \Big) {F_{\rm TT}(Q_1^2,Q_2^2) \over 2X}  + {Q_2 M \over X} F_{\rm TL}(Q_1^2,Q_2^2) \Big \} \, .  \\
{1 \over 4 \pi \alpha_{\rm em}} {\cal M}(\pm) &=&  i [\bq_1,\bq_2]
 (\bq_2 \cdot \bE^*(\pm)) {M F_{\rm TT}(Q_1^2, Q_2^2) \over X}
+ i [\bq_1,\bq_2]  (\bP \cdot \bE^*(\pm)) {Q_2 \over X} F_{\rm TL}(Q_1^2,Q_2^2) \nonumber \\
&+& {M^2 + \bP^2 \over 2 X} \Big( i [\bE^*(\pm),\bq_2] Q_1  
F_{\rm LT}(Q_1^2,Q_2^2) + i [\bE^*(\pm),\bq_1] Q_2 F_{\rm TL}(Q_1^2,Q_2^2) \Big) \, .  \nonumber
\end{eqnarray}   
For the extraction of form factors from our amplitude, it is convenient to rewrite the transverse amplitude using the identities of Eq.~(\ref{eq:vector_identity}).
We can then represent our helicity amplitudes as follows:
\begin{eqnarray}
 {X \over 4 \pi \alpha_{\rm em}}{\cal M}(0) &=& -i [\bq_1,\bq_2] \Big\{ \half \Big(M^2 +Q_1^2 - Q_2^2 \Big) F_{\rm TT}(Q_1^2,Q_2^2) + M Q_2
F_{\rm TL}(Q_1^2,Q_2^2) \Big \} \, ,  \\
{X \over 4 \pi \alpha_{\rm em}} {\cal M}(\pm) &=& i[\bq_1,\bE^*] \Big( Q_1^2 M F_{\rm TT}(Q_1^2,Q_2^2) - \half (M^2 + Q_1^2 - Q_2^2) Q_2 F_{\rm TL}(Q_1^2,Q_2^2)  \Big ) \nonumber \\
&& - i[\bq_2,\bE^*] \Big( Q_1^2 Q_2 F_{\rm TL}(Q_1^2,Q_2^2) + 
\half (M^2 + Q_1^2 + Q_2^2) Q_1 F_{\rm LT}(Q_1^2,Q_2^2) \Big) \nonumber \\ 
&& - [\bq_2,\bE^*] (\bq_1 \cdot \bq_2) \Big( M F_{\rm TT}(Q_1^2,Q_2^2) + Q_2 F_{\rm TL}(Q_1^2,Q_2^2) + Q_1 F_{\rm LT}(Q_1^2,Q_2^2) \Big)\, . \nonumber
\label{eq:calM}
\end{eqnarray}
It is useful to write down symmetric ($F_S$) and antisymmetric ($F_A$) combinations of form factors,
\begin{eqnarray}
F_S (Q_1^2,Q_2^2) &\equiv &  Q_2 F_{\rm TL} (Q_1^2,Q_2^2) - Q_1 F_{\rm LT}(Q_1^2,Q_2^2) \, ,\nonumber \\
F_A (Q_1^2,Q_2^2) &\equiv &  Q_2 F_{\rm TL} (Q_1^2,Q_2^2) + Q_1 F_{\rm LT}(Q_1^2,Q_2^2) \, .
\end{eqnarray}
Now we wish to relate the form factors $\Phi_i$ of our LF amplitude to the invariant form factors of the covariant amplitude in Eq.~(\ref{eq:formfactors}). To this end, it is instructive to turn to the helicity amplitudes corresponding to \emph{linear} polarizations of the meson. These can then be expanded in a Fourier series in $\cos n\phi, \sin n\phi$, where $\phi$ is the angle between the photon transverse momenta $\bq_1,\bq_2$. Here, we use the following frame: the longitudinal polarization defines the $z$-direction, and the $x$-direction is taken along the meson transverse momentum $\bP = \bq_1 + \bq_2$. For a covariant definition of this particular helicity frame, see Appendix~\ref{sec:linear_pol}. We can now project the amplitude onto the linear polarizations and obtain the following results:
\begin{eqnarray}
(-i){\cal T}_z &=& Q_1 Q_2 \sin \phi \, \Phi_0 \ , \nonumber \\
(-i) |\bP| \,  {\cal T}_x &=&  Q_1 Q_2 \sin \phi \, \Big(\Phi_1 - \Phi_2 \Big) \, , \nonumber \\
(-i) |\bP| \, {\cal T}_y
&=& Q_1^2 \Phi_1 + Q_2^2 \Phi_2 + Q_1 Q_2 \cos \phi \Big( \Phi_1 + \Phi_2  \Big) \, . 
\label{eq:linear_T}
\end{eqnarray}
We have collected rather lengthy expressions for our covariant amplitude in Appendix~\ref{sec:linear_pol}. Let us now make some observations on these results. Firstly, we see that in the amplitudes of Eq.~(\ref{eq:linear_T}), there are no terms proportional to $\cos 2 \phi$ or $\sin 2 \phi$. These do appear, however, in the decomposition of the general invariant amplitude ${\cal M}$ -- they obviously are related to the last term in Eq.~(\ref{eq:calM}). These terms are all multiplied by the combination,
\begin{eqnarray}
f_A(Q_1^2,Q_2^2) \equiv M F_{\rm TT}(Q_1^2,Q_2^2) + F_A(Q_1^2,Q_2^2) \, ,
\end{eqnarray}
which, for consistency with Eq.~(\ref{eq:linear_T}), has to vanish identically, i.e. 
\begin{eqnarray}
f_A(Q_1^2,Q_2^2) \equiv 0\, .
\label{eq:cancellation}
\end{eqnarray}
This appears to be a peculiar identity related to the $\gamma^* \gamma^* \to Q \bar Q$ helicity amplitudes.
Consequently, the number of independent form factors is reduced to two, say, $F_A, F_S$. In effect, the form factor for two transverse photons is entirely determined through a mixture of longitudinal/transverse ($F_{\rm LT}, F_{\rm TL}$) form factors.

Interestingly, after exploiting the cancellation of Eq.~(\ref{eq:cancellation}), all our amplitudes have simple symmetry properties, namely:
\begin{eqnarray}
{\cal M}_z &=& {-i Q_1 Q_2 \over 2 X} \cdot \sin \phi \cdot f_S(Q_1^2, Q_2^2) \,, \nonumber \\
|\bP| {\cal M}_x &=& {-i Q_1 Q_2 \over 2X} \cdot \sin \phi \cdot M f_S(Q_1^2,Q_2^2) \,, \nonumber \\
|\bP| {\cal M}_y &=& {i \over 2X} 
\Big\{ \Big( 2 Q_1^2 Q_2^2  + \nu (Q_1^2 + Q_2^2) \Big)  F_A(Q_1^2,Q_2^2) + \nu (Q_1^2 - Q_2^2) F_S(Q_1^2,Q_2^2) 
 \nonumber \\
&+& Q_1 Q_2 \cos \phi \Big[
(2 \nu + Q_1^2 + Q_2^2) F_A(Q_1^2,Q_2^2) + (Q_1^2 - Q_2^2) F_S(Q_1^2,Q_2^2)  \Big] \Big\} \, .
\end{eqnarray}
Here, we defined
\begin{eqnarray}
\nu \equiv q_1 \cdot q_2 = \half (M^2 + Q_1^2 + Q_2^2 ) \, ,
\end{eqnarray}
as well as
\begin{eqnarray}
f_S(Q_1^2,Q_2^2) &\equiv & (Q_1^2 - Q_2^2) 
F_{\rm TT}(Q_1^2,Q_2^2) + M F_S(Q_1^2,Q_2^2) \nonumber \\
&=& - {Q_1^2 - Q_2^2 \over M} F_A(Q_1^2,Q_2^2) + M F_S(Q_1^2, Q_2^2) \, .
\end{eqnarray}

Let us now proceed and try to extract form factors $F_A, F_S$. To this end, we chose the $y$-component of the amplitude, i.e. the linear polarization orthogonal to the ``production plane''. As a short-hand notation, we define the form factors
\begin{eqnarray}
G_1(Q_1^2,Q_2^2) &\equiv & Q_1^2 \Phi_1(Q_1^2,Q_2^2) + Q_2^2 \Phi_2(Q_1^2,Q_2^2) \nonumber \\
&=& \half (Q_1^2 + Q_2^2) \Phi_A(Q_1^2,Q_2^2)  
+ \half (Q_1^2 - Q_2^2) \Phi_S(Q_1^2,Q_2^2) \, ,
\nonumber \\
G_2(Q_1^2,Q_2^2) &\equiv & \Phi_1(Q_1^2,Q_2^2)+ \Phi_2(Q_1^2,Q_2^2) = \Phi_A(Q_1^2,Q_2^2)   \, , 
\end{eqnarray}
where
\begin{eqnarray}
\Phi_S \equiv \Phi_1 - \Phi_2 \, , \qquad
\Phi_A \equiv  \Phi_1 + \Phi_2 \, , 
\label{eq:Phi_S_Phi_A}
\end{eqnarray}
are symmetric and antisymmetric in $Q_1^2 \leftrightarrow Q_2^2$ combinations. We then derive a system of linear equations:
\begin{eqnarray}
e_f^2 \sqrt{N_c} \, \begin{pmatrix}
G_1 \\ G_2
\end{pmatrix}
= {1 \over 2X} \begin{pmatrix}
2 Q_1^2 Q_2^2 + \nu (Q_1^2 + Q_2^2) & \nu (Q_1^2 - Q_2^2) \\
2 \nu + Q_1^2 + Q_2^2 & Q_1^2 - Q_2^2
\end{pmatrix}
\begin{pmatrix}
F_A \\ F_S 
\end{pmatrix} \,.
\end{eqnarray}
Here, the determinant of the $2\times 2$ matrix is $\det(\dots) = - 2X (Q_1^2 - Q_2^2)$ so that we can invert the matrix to obtain 
\begin{eqnarray}
\begin{pmatrix}
F_A \\ F_S
\end{pmatrix}
= { - e_f^2 \sqrt{N_c} \over Q_1^2 - Q_2^2 }
\begin{pmatrix}
Q_1^2 - Q_2^2 & - \nu (Q_1^2 - Q_2^2) \\
- (2 \nu + Q_1^2 + Q_2^2) & 2 Q_1^2 Q_2^2 + \nu (Q_1^2+Q_2^2) 
\end{pmatrix}
\begin{pmatrix}
G_1 \\ G_2
\end{pmatrix} \,.
\end{eqnarray}
This finally gives us
\begin{eqnarray}
F_A &=& e_f^2 \sqrt{N_c} \Big\{ ( \nu - Q_1^2) \Phi_1  + (\nu - Q_2^2) \Phi_2 \Big\} \nonumber \\
&=& {e_f^2 \sqrt{N_c} \over 2}  \Big\{  M^2 \Phi_A -  (Q_1^2 - Q_2^2) \Phi_S \Big\} \, ,
\end{eqnarray}
and
\begin{eqnarray}
F_S &=&  e_f^2 \sqrt{N_c} \Big\{ (\nu + Q_1^2) \Phi_1 - (\nu + Q_2^2) \Phi_2 \Big\} \nonumber \\
&=& {e_f^2 \sqrt{N_c} \over 2} \Big\{ (M^2 + 2 Q_1^2 + 2 Q_2^2) \Phi_S + (Q_1^2 - Q_2^2) \Phi_A \Big\} \, ,
\end{eqnarray}
or, finally, the original form factors $F_{\rm TL},F_{\rm LT}$:
\begin{eqnarray}
Q_1 F_{\rm LT} &=& {e_f^2 \sqrt{N_c} \over 2} \Big\{ (\nu - Q_1^2) \Phi_A - (\nu + Q_1^2) \Phi_S \Big\} \, , \nonumber \\
Q_2 F_{\rm TL} &=& {e_f^2 \sqrt{N_c} \over 2} \Big\{ (\nu - Q_2^2) \Phi_A + (\nu + Q_2^2) \Phi_S \Big\} \, , 
\end{eqnarray}
and
\begin{eqnarray}
F_{\rm TT} = - {1 \over M} \Big\{ Q_1 F_{\rm LT} + Q_2 F_{\rm TL} \Big\} \, .
\end{eqnarray}

A final comment on the determination of $F_{\rm TT},F_{\rm TL},F_{\rm LT}$ is in order. Evidently, we have at our disposal three helicity amplitudes, and we start out with three form factors. However, as it turns out, a cancellation that relates $F_{\rm TT}$ with $F_{\rm TL}$ and $F_{\rm LT}$ makes our system over determined. In fact, we could have obtained our form factors $F_A,F_S$ just as well from the comparison of ${\cal M}_x$ and ${\cal M}_y$ or from ${\cal M}_z$ and one of the transverse polarizations. If we stay within the transverse polarizations, there is no ambiguity, and the expressions in term of $\Phi_{1,2}$ will be the same. However, involving the longitudinal polarization involves the form factor $\Phi_0$, which is not related in any obvious way to $\Phi_1, \Phi_2$. This issue is evidently related to the problems with rotational symmetry in LF quantization. We believe that it is an artefact of the treatment of the bound state as a pure $Q \bar Q$ state, however, a detailed discussion of the latter goes beyond the scope of this work.

Before we come to our full numerical results including the integrals over the quarkonium wave function, we present the NRQCD limits of the invariant form factors. They are all related to the derivative of the rest frame WF at the origin, $R'(0)$, as follows:
\begin{eqnarray}
F_{\rm TT}(Q_1^2,Q_2^2) &=& 2  e_f^2 \sqrt{6 N_c \over  \pi M^3} R'(0) \, {Q_1^2 - Q_2^2 \over \nu } \, , \nonumber \\
F_{\rm LT}(Q_1^2,Q_2^2) &=& - 2 e_f^2 \sqrt{6 N_c \over \pi M} R'(0) {(\nu + Q_2^2 ) Q_1\over \nu^2} \, , \nonumber \\
F_{\rm TL}(Q_1^2,Q_2^2) &=& 2 e_f^2 \sqrt{6 N_c \over  \pi M} R'(0) {(\nu + Q_1^2 ) Q_2\over \nu^2} \, ,
\end{eqnarray}
where, within the accuracy of the NRQCD approximation, one should use $M = 2m_Q$.

\section{Form factor visualization and observables}
\label{sec:ff_results}

We now come to the presentation of our numerical results. 
First we will discuss our results for the form factors $\Phi_{A,S}$ of the light front amplitude from which all invariant form factors are calculated. 

Then, we will recall the main observables which can be extracted in $e^+ e^-$ reactions and how they are related to the invariant form factors.

All our results are obtained for the $\chi_{c1}$ meson. Below we use the mass $M=3.410 \, \rm{GeV}$. The light front wave functions were obtained from the radial wave functions in the $1P$ partial wave. More details on the WF and the $c \bar c$ potentials used can be found in Ref.~\cite{Babiarz:2020jkh}.

\subsection{Form factors of the LF amplitude}

Let us first discuss the form factors $\Phi_{0,1,2}$ introduced in Sec.~\ref{sec:LC-formulation}. These are the primary results of our numerical calculations. First let us have a look at the combinations $\Phi_S$ and $\Phi_A$ of Eq.~(\ref{eq:Phi_S_Phi_A}). As we have shown on general grounds, these form factors have good symmetry properties under the interchange $Q_1^2 \leftrightarrow Q_2^2$ -- they are respectively antisymmetric ($\Phi_A$) and symmetric ($\Phi_S$). These symmetry properties are not obvious from the expressions Eq.~(\ref{eq:Phi_1_Phi_2}), and hence are a good check on the numerics. To this end, we introduce the average of virtualities, $\overline{Q^2}$, and asymmetry, $\omega$,
\begin{eqnarray}
\overline{Q^2} = {Q^2_1 + Q^2_2 \over 2}\,, \qquad \omega = {Q^2_2 -Q^2_1 \over Q^2_1 + Q^2_2} \,.
\end{eqnarray}
\begin{figure}[h!]
    \centering
    \includegraphics[width = 0.32\textwidth]{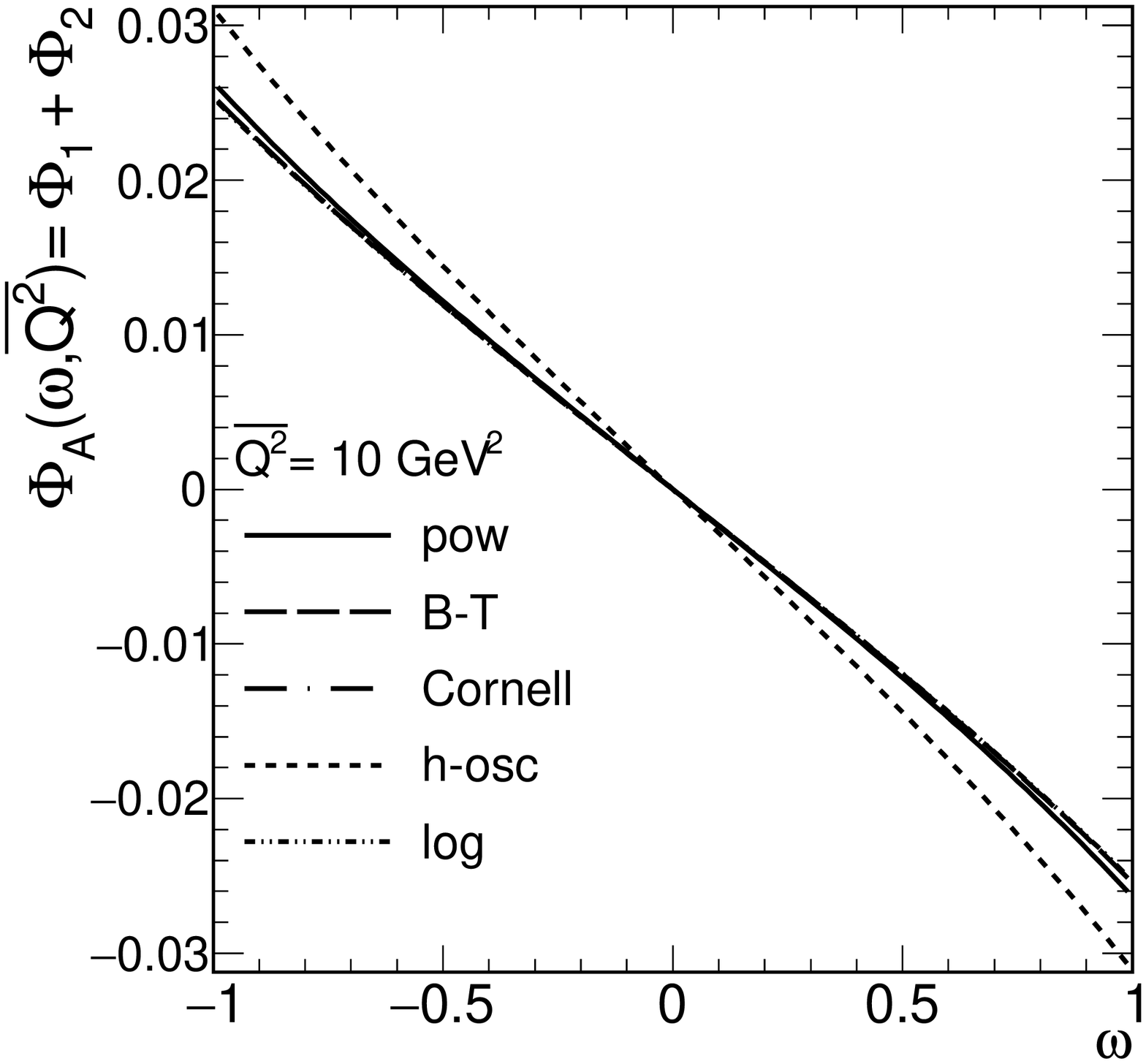}
    \includegraphics[width = 0.32\textwidth]{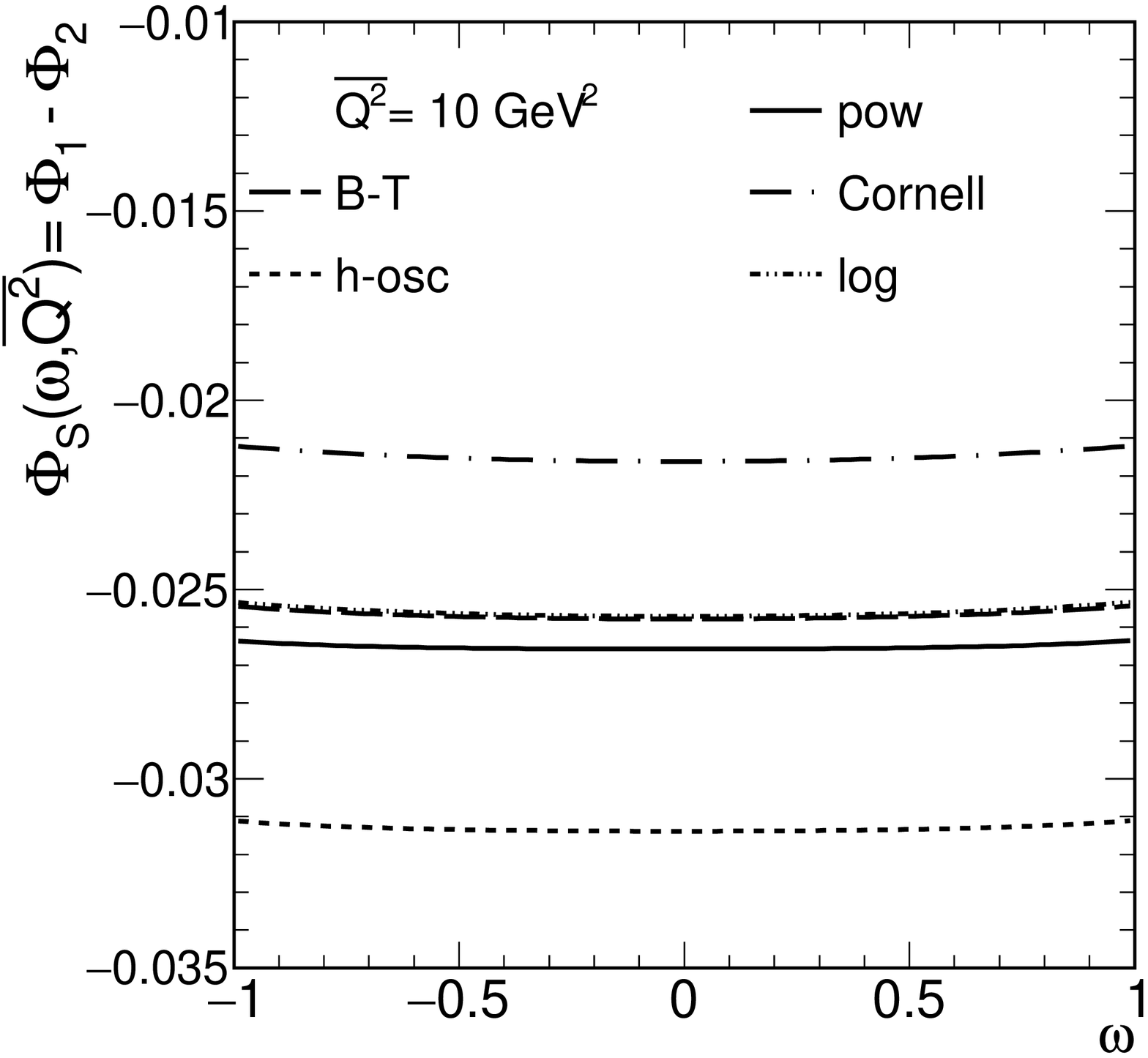}
    \includegraphics[width =0.32\textwidth]{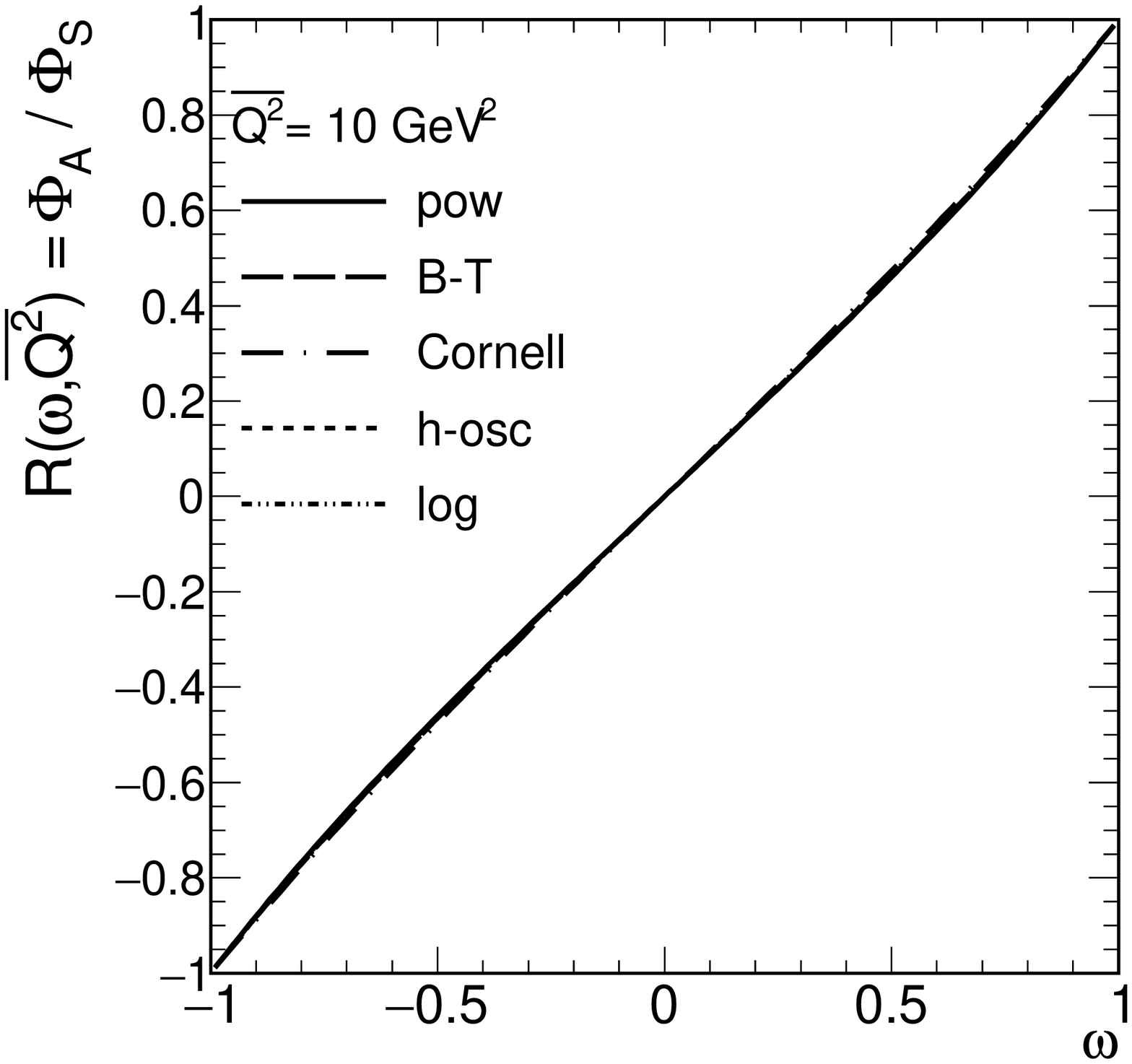}
\caption{Antisymmetric $\Phi_A$ and symmetric $\Phi_S$ form factors for $\gamma^* \gamma^* \to \chi_c (1^{++})$ as a function of the asymmetry of photon virtualities (left and middle panels)
 and the ratio of the $\Phi_A/\Phi_S$ form factors again as a function of asymmetry of photon virtualities (right panel).}
    \label{fig:PhiA_PhiS}
\end{figure}
\begin{figure}
    \centering
    \includegraphics[width = 0.32\textwidth]{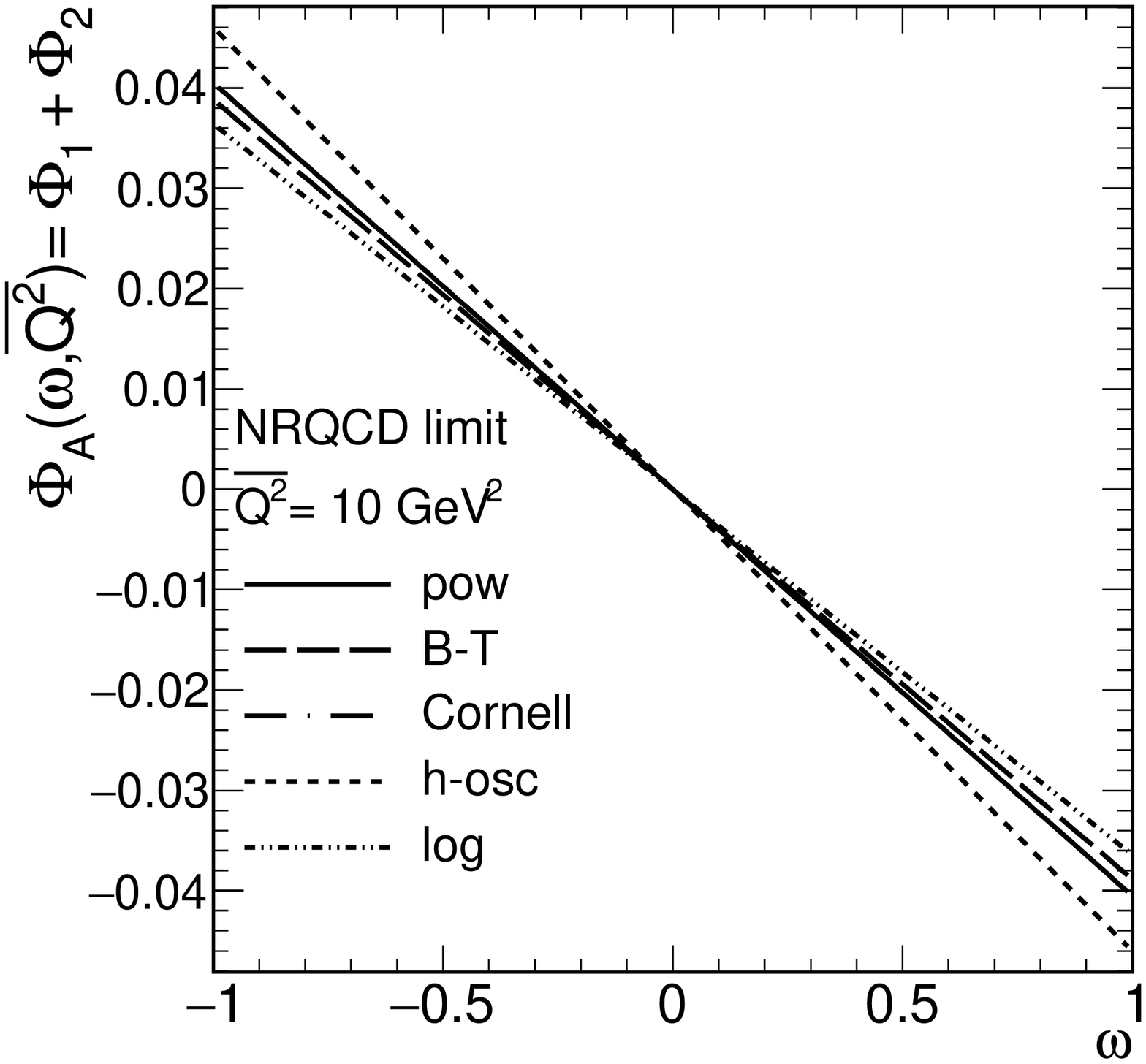}
    \includegraphics[width =0.32\textwidth]{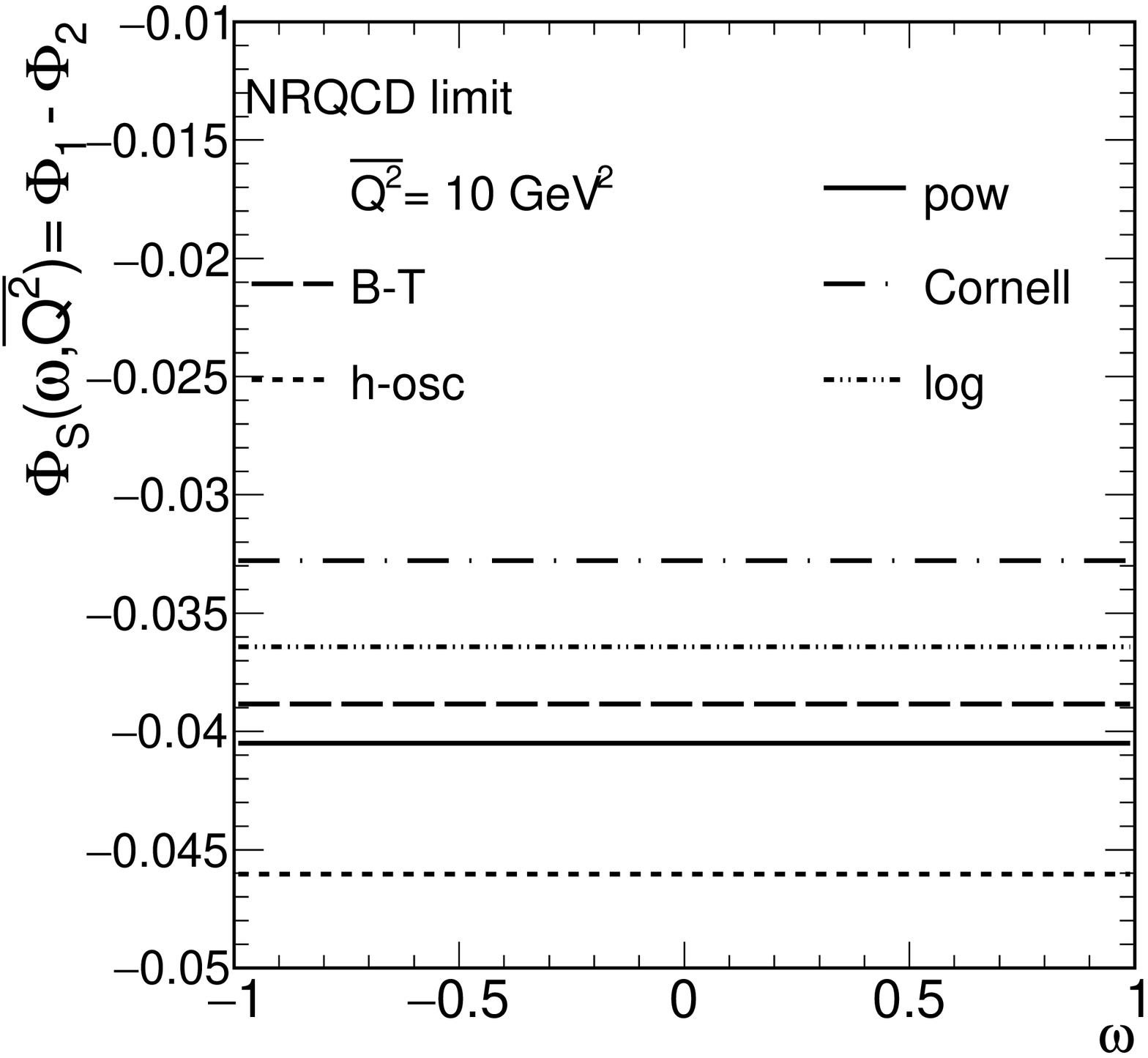}
    \includegraphics[width = 0.32\textwidth]{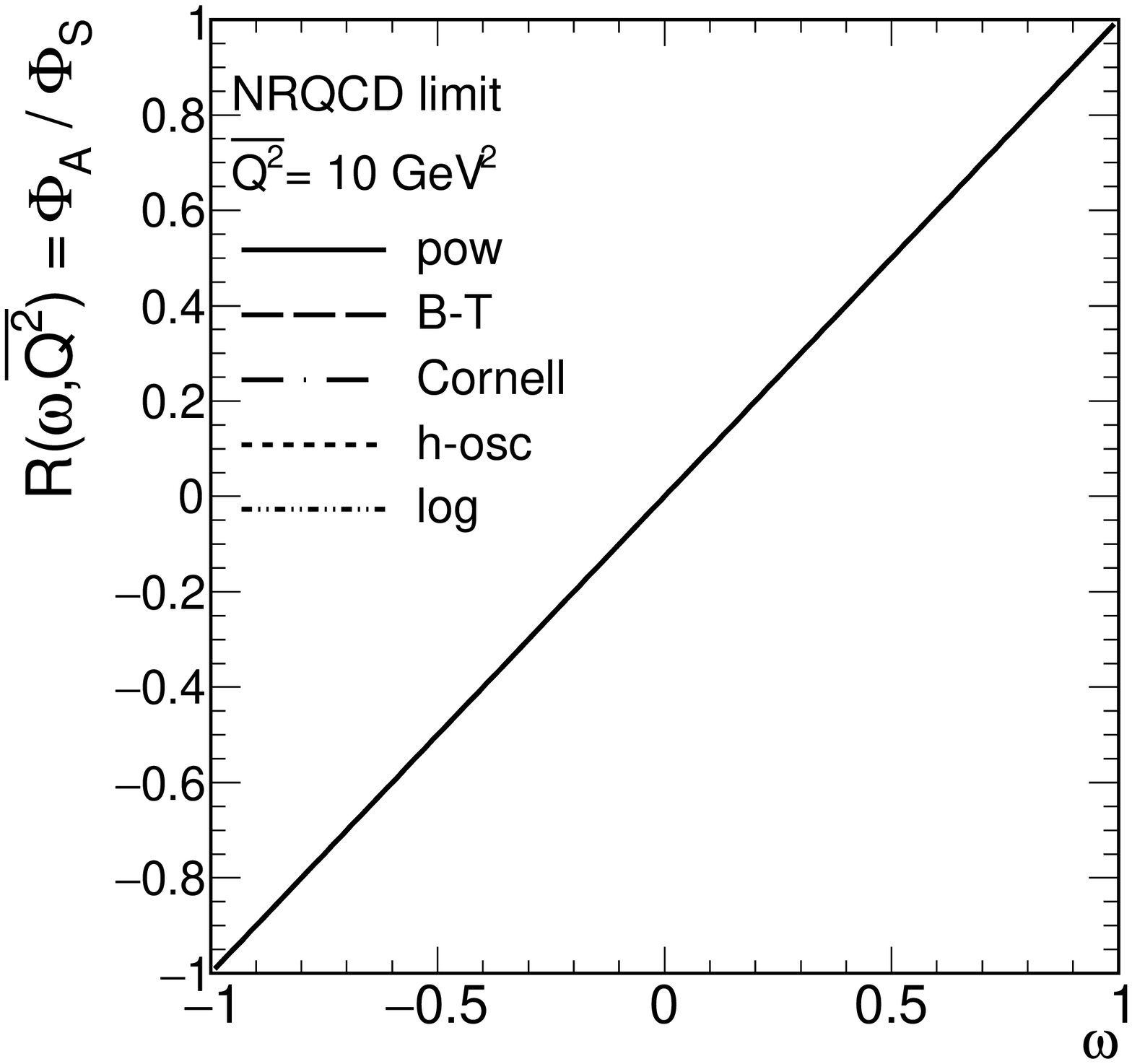}
    \caption{Antisymmetric $\Phi_A$ and symmetric $\Phi_S$ form factors in the nonrelativistic limit with the specific value of the quark mass $m_c$ and the first derivative $R'(0)$ depending on each potential model, see Tab.~\ref{tab:reduced_width}.}
    \label{fig:PhiA_PhiS_NR}
\end{figure}
\begin{figure}[h!]
    \centering
    \includegraphics[width= 0.33\textwidth]{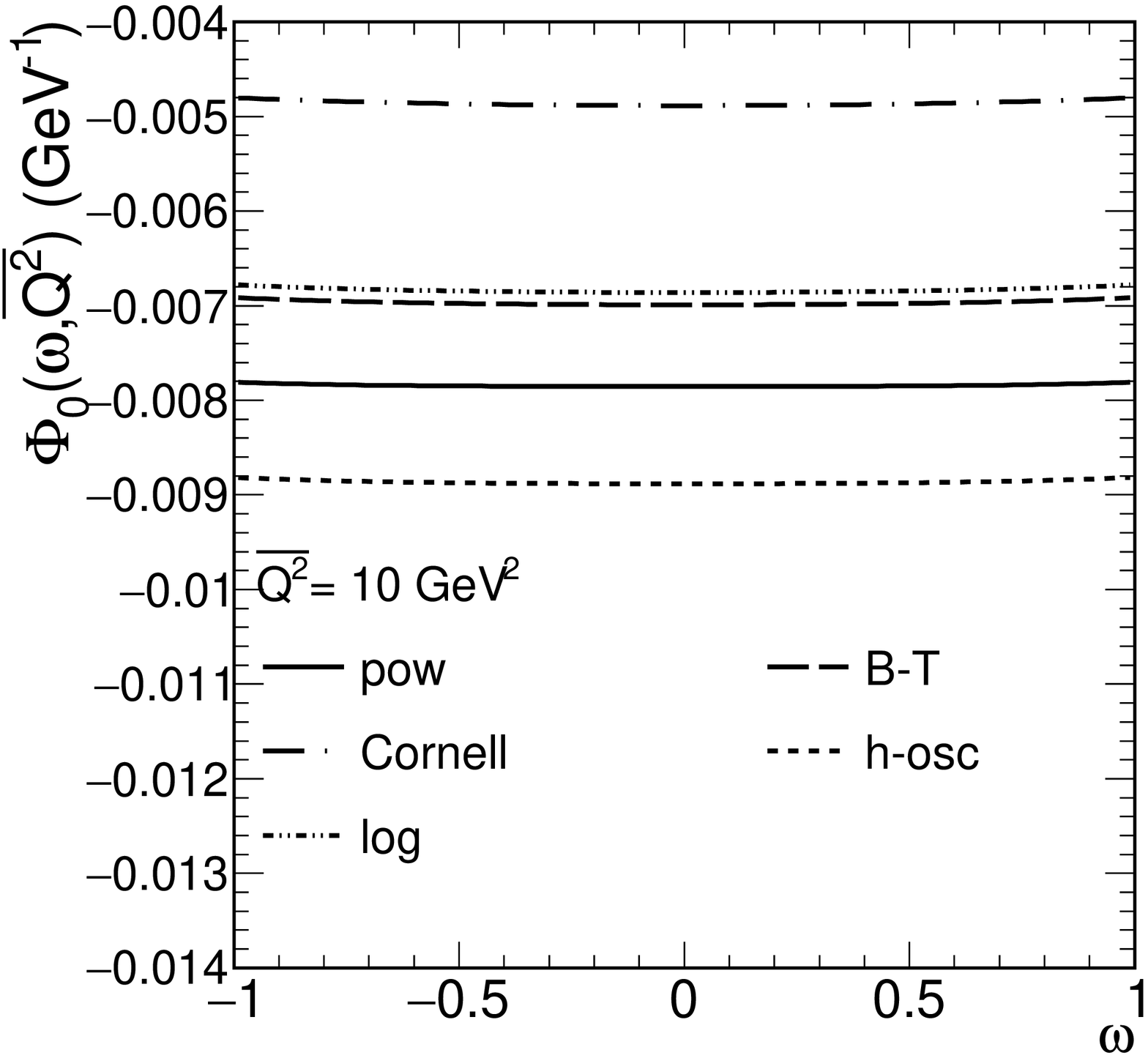}
    \includegraphics[width= 0.33\textwidth]{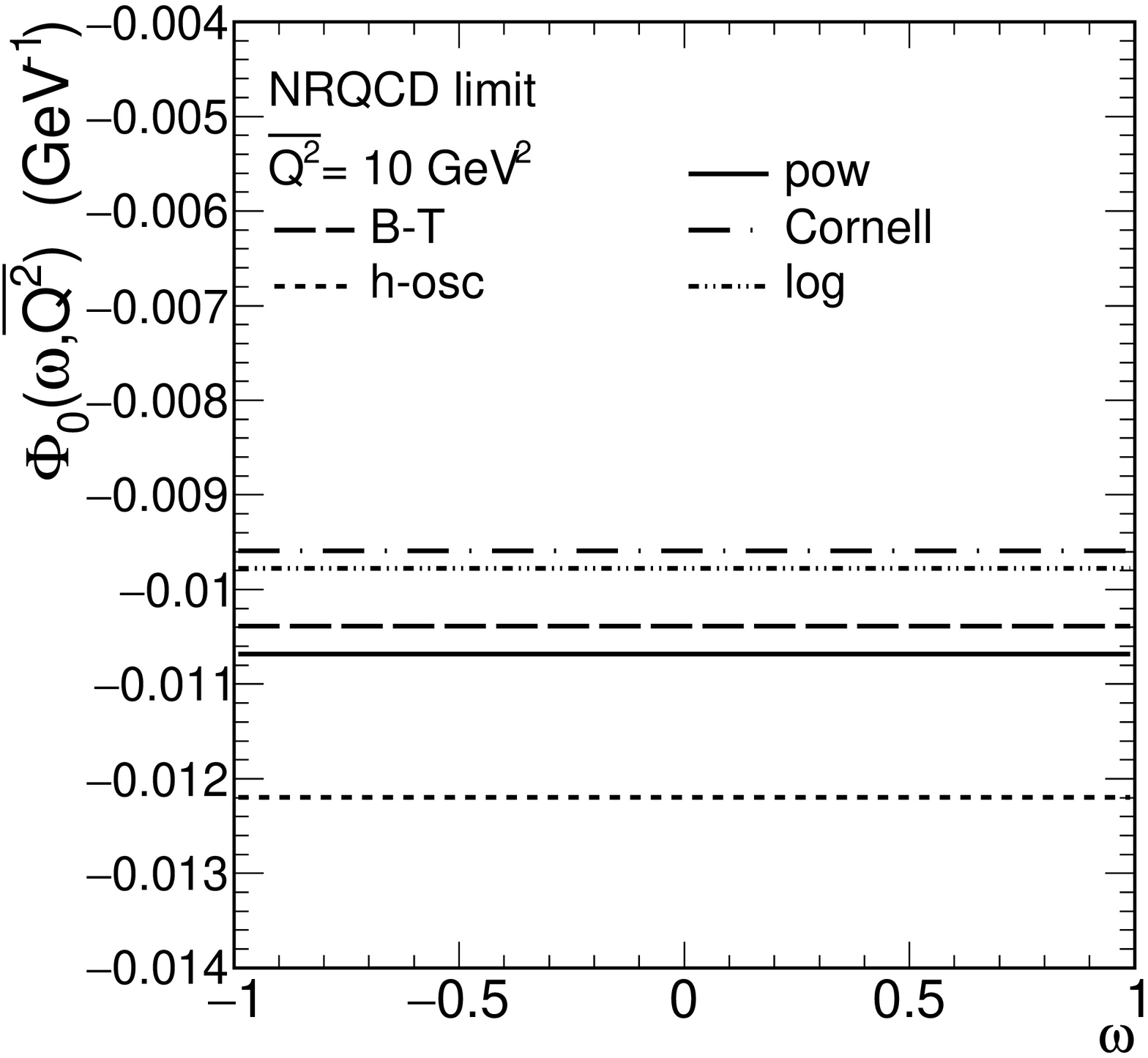}
    \caption{Invariant form factor $\Phi_0$ as a function of asymmetry parameter with fixed $\bar{Q^2} = 10\,{\rm GeV}^2$,
    on the l.h.s. within LFWF approach, and on the r.h.s. in the nonrelativistic limit.}
    \label{fig:Phi0}
\end{figure}

In Fig.~\ref{fig:PhiA_PhiS} we present the $\Phi_A$ and $\Phi_S$ form factors as a function of asymmetry parameter $\omega$ for fixed value of $\overline{Q^2}$ = 10 GeV$^2$. We observe that the expected symmetry behaviour is well borne out by our numerical results. In Fig.~\ref{fig:PhiA_PhiS_NR} we show the corresponding NRQCD results given by Eq.~(\ref{eq:Phi_NRQCD}). In the NRQCD limit the relation $\Phi_A = \omega \Phi_S$ holds true, as can be seen from the ratio plotted in the third panel of Fig~\ref{fig:PhiA_PhiS_NR}. There is no reason for this relation to remain satisfied also in the full LFWF approach, but the rightmost panel of Fig.~\ref{fig:PhiA_PhiS} shows that it is a surprisingly good approximation, independently of the potential used. 

For completeness, in Fig.~\ref{fig:Phi0} we show the form factor $\Phi_0$ of the amplitude in Eqs.~(\ref{eq:def_Phi_0},\ref{eq:Integrals_Phi_0}) as a function of $\omega$. In the NRQCD limit $\Phi_0$ does not depend on $\omega$, while the full result shows a very weak dependence on $\omega$.

\subsection{Form factors for two virtual photons}

\begin{figure}[h!]
    \centering
    \includegraphics[width = 0.49\textwidth]{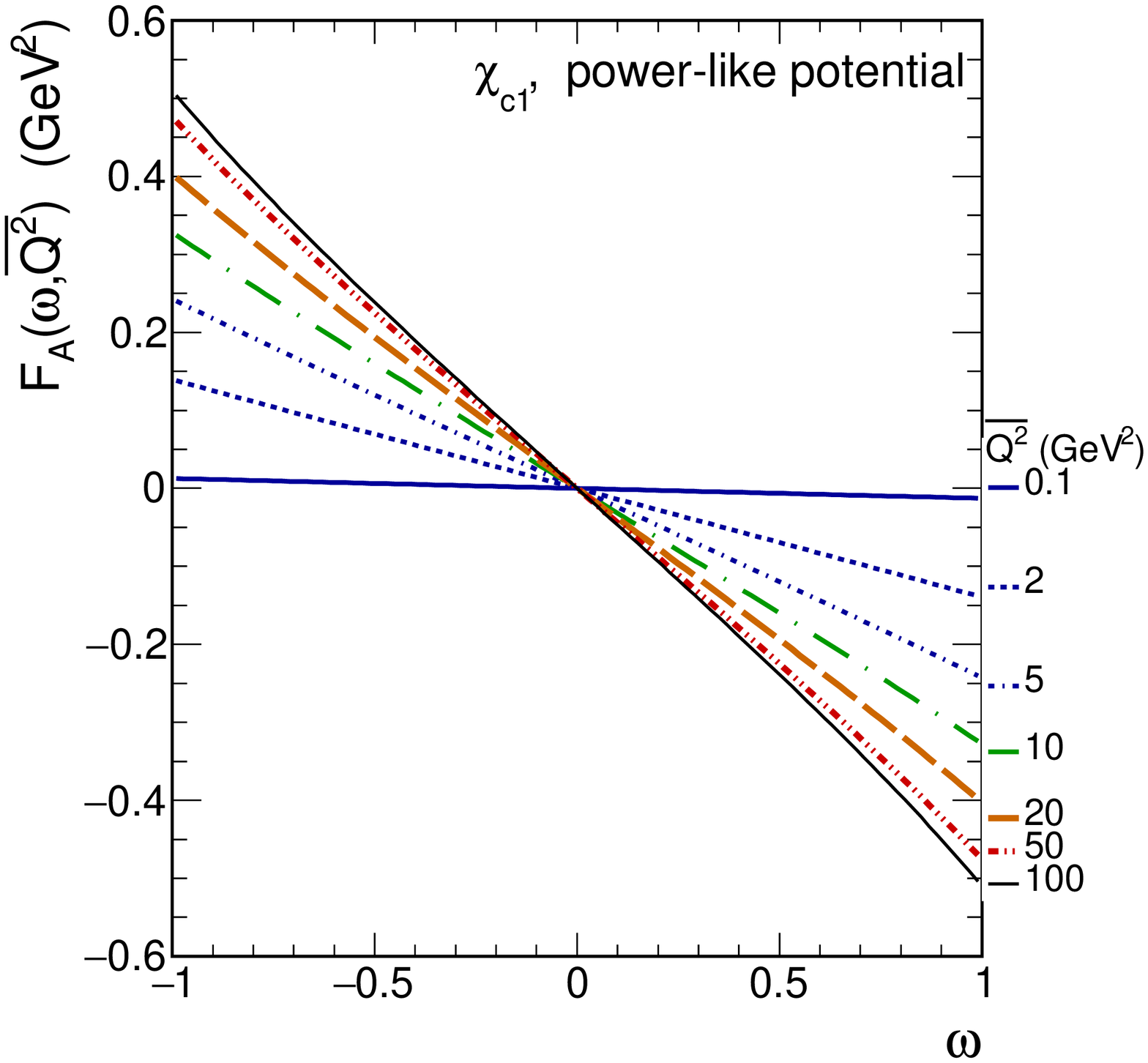}
    \includegraphics[width = 0.49 \textwidth]{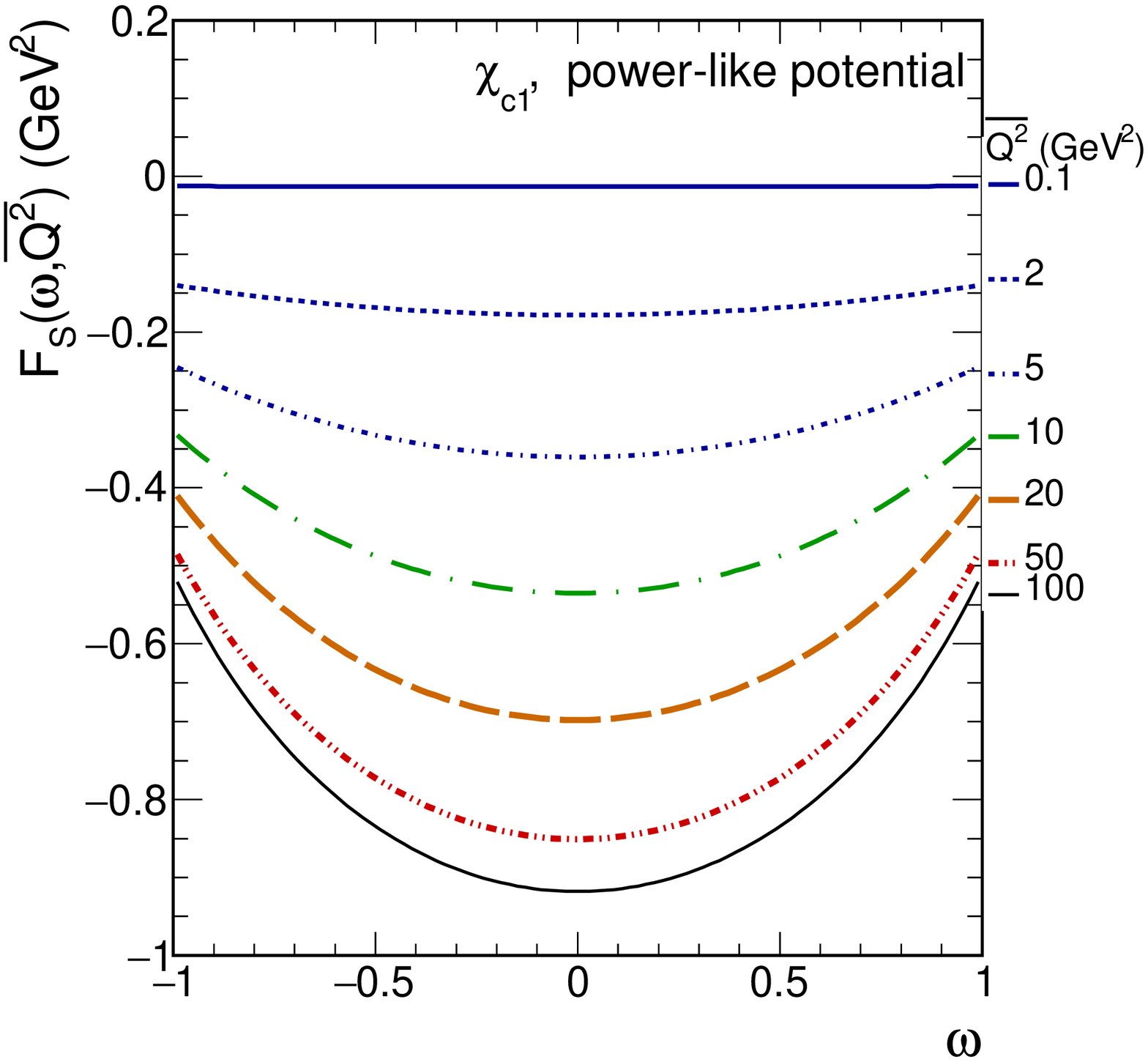}
    \caption{Antisymmetric $F_A$ and symmetric $F_S$ form factors as functions of asymmetry parameter 
$\omega$ for several values of $\overline{Q^2}$ in the range $(0.1 - 100)\, {\rm GeV}^{2}$ for the power-like potential.}
    \label{fig:FA_FS_omega_Q2ave}
\end{figure}
\begin{figure}
    \centering
    \includegraphics[width = 0.49\textwidth]{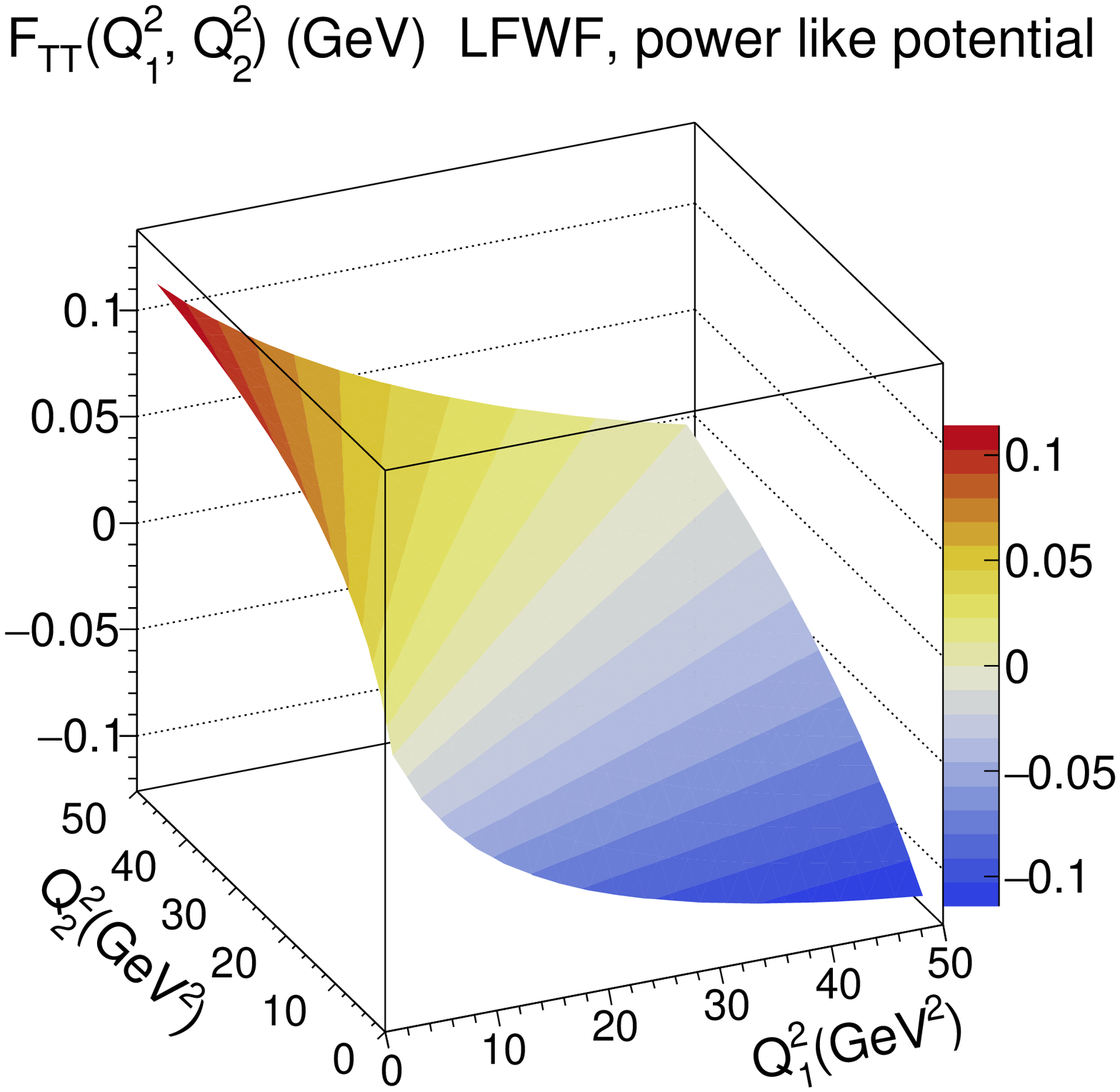}
    \includegraphics[width =0.49\textwidth]{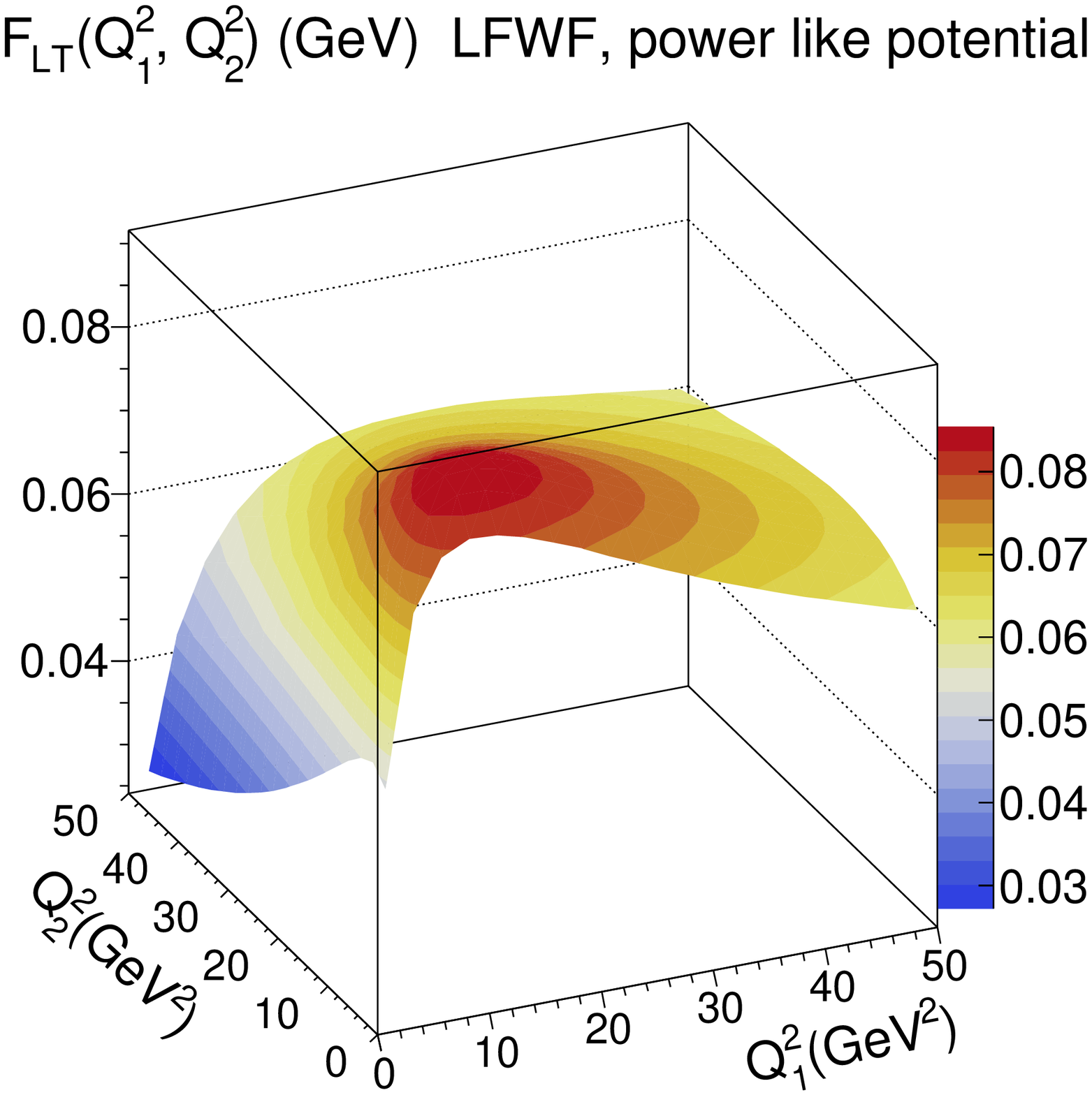}
    \caption{
    Dependence of form factors $F_{\rm TT}(Q^2_1,Q^2_2)$ and $F_{\rm LT}(Q^2_1,Q^2_2)$ on the two photon virtualities. Here we used the LFWF obtained from the power-like potential model.
    }
    \label{fig:FTT_FLT_3D}
\end{figure}

In Fig.~\ref{fig:FA_FS_omega_Q2ave} we show the antisymmetric $F_A$ and symmetric $F_S$ form factors as functions of photon virtuality asymmetry for different values of $\overline{Q^2}$. These plots were obtained for the power-law potential. Both form factors decrease in magnitude  when $\overline{Q^2} \to 0$. In Fig.~\ref{fig:FTT_FLT_3D} we show maps of helicity form factors $F_{\rm TT}$ and $F_{\rm LT}$ in the $(Q_1^2, Q_2^2)$-plane. The form factor $F_{\rm TT}$ changes its sign when crossing the $Q_1^2 = Q_2^2$ line, while $F_{\rm LT}$ has no symmetry properties. Also these plots were obtained for the power-law potential. Qualitatively, the plots for different potentials look similar. 

\subsection{Form factors for one real and one virtual photon}

For a quantitative representation of the differences between the underlying $c \bar c$-potentials it is more convenient to turn to one-dimensional form factors with only one off--shell photon, which we wish to discuss now.

\begin{figure}[h!]
    \centering
    \includegraphics[width = 0.32\textwidth]{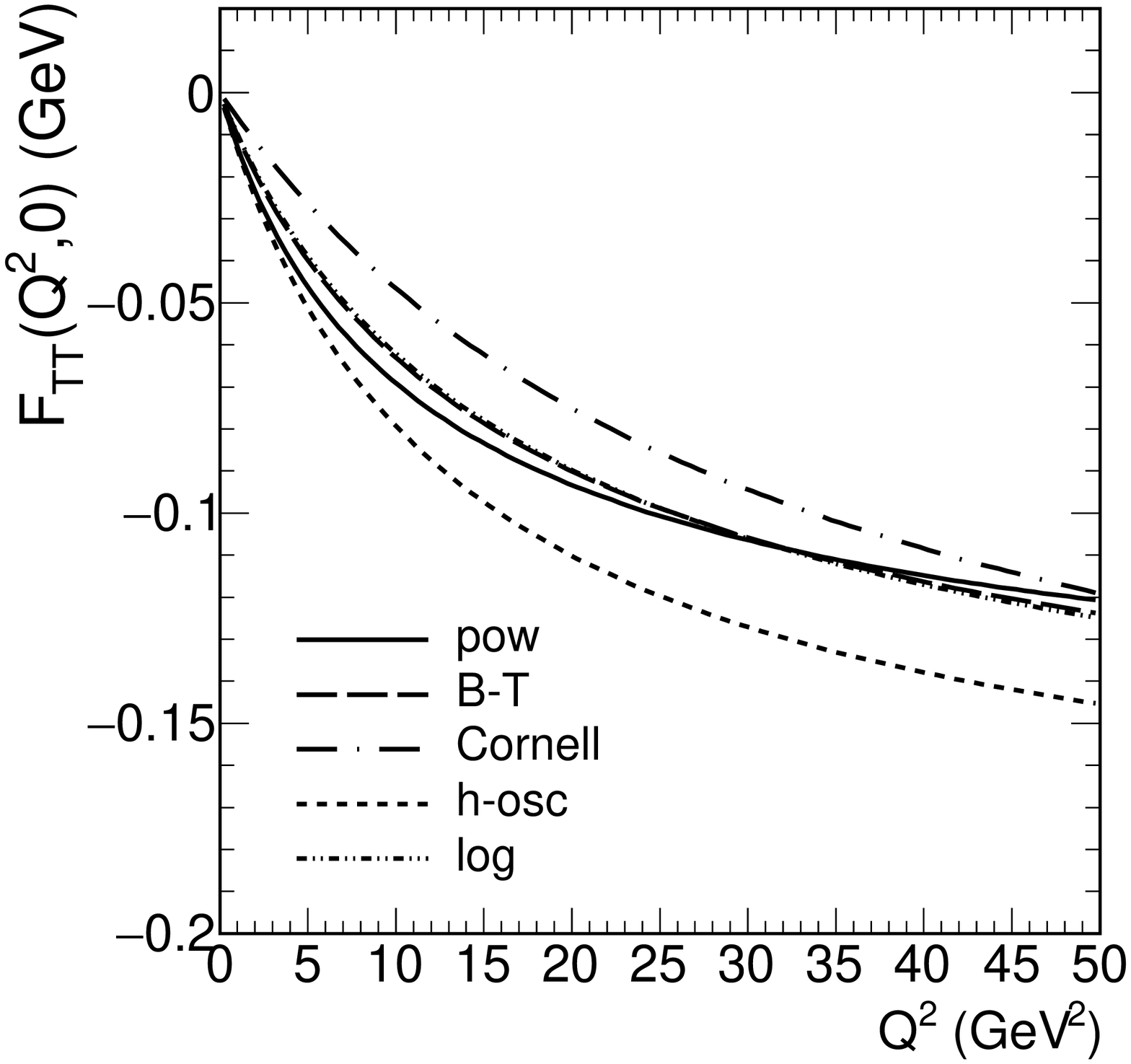}
    \includegraphics[width = 0.32\textwidth]{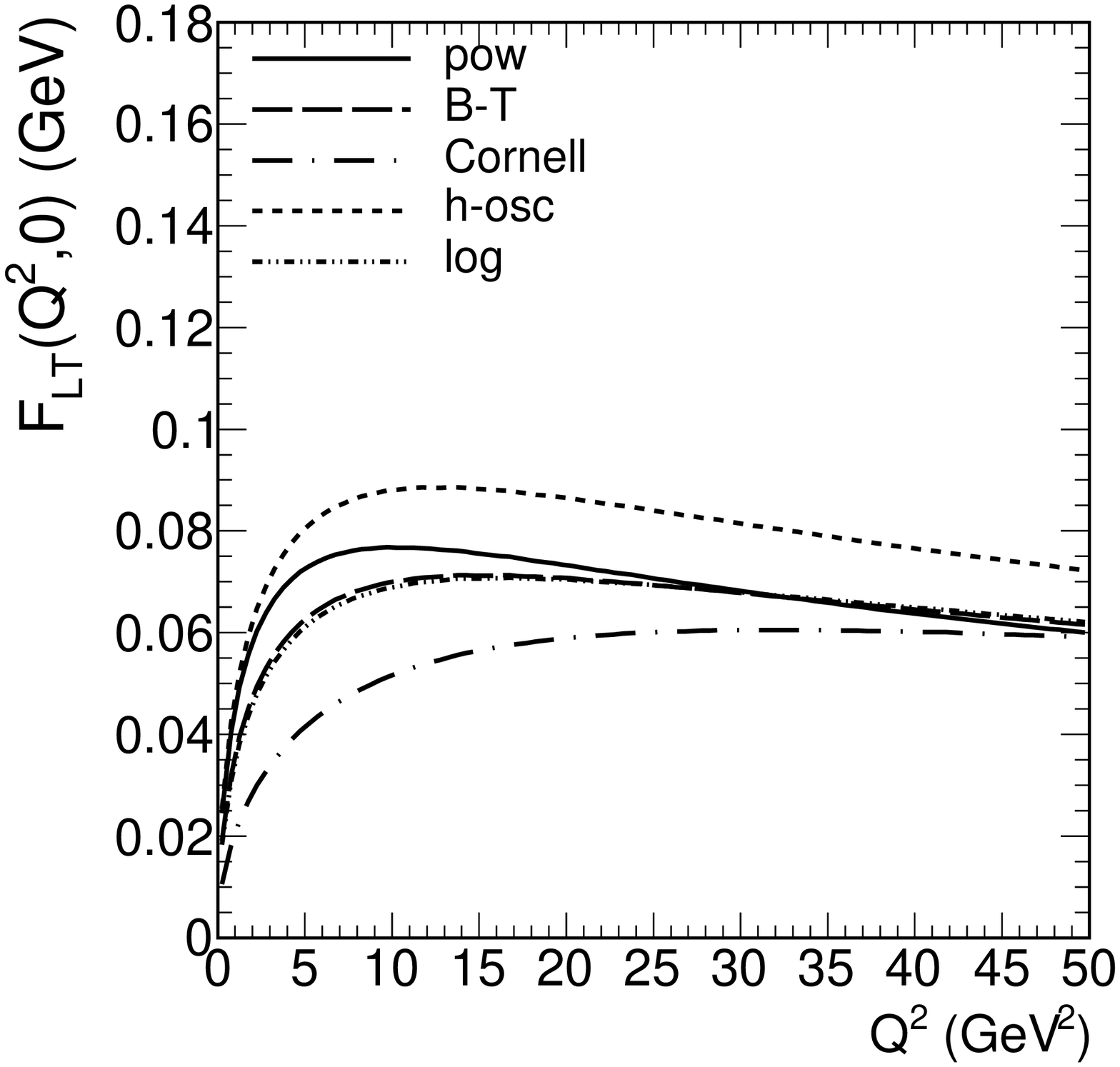}
    \includegraphics[width = 0.32\textwidth]{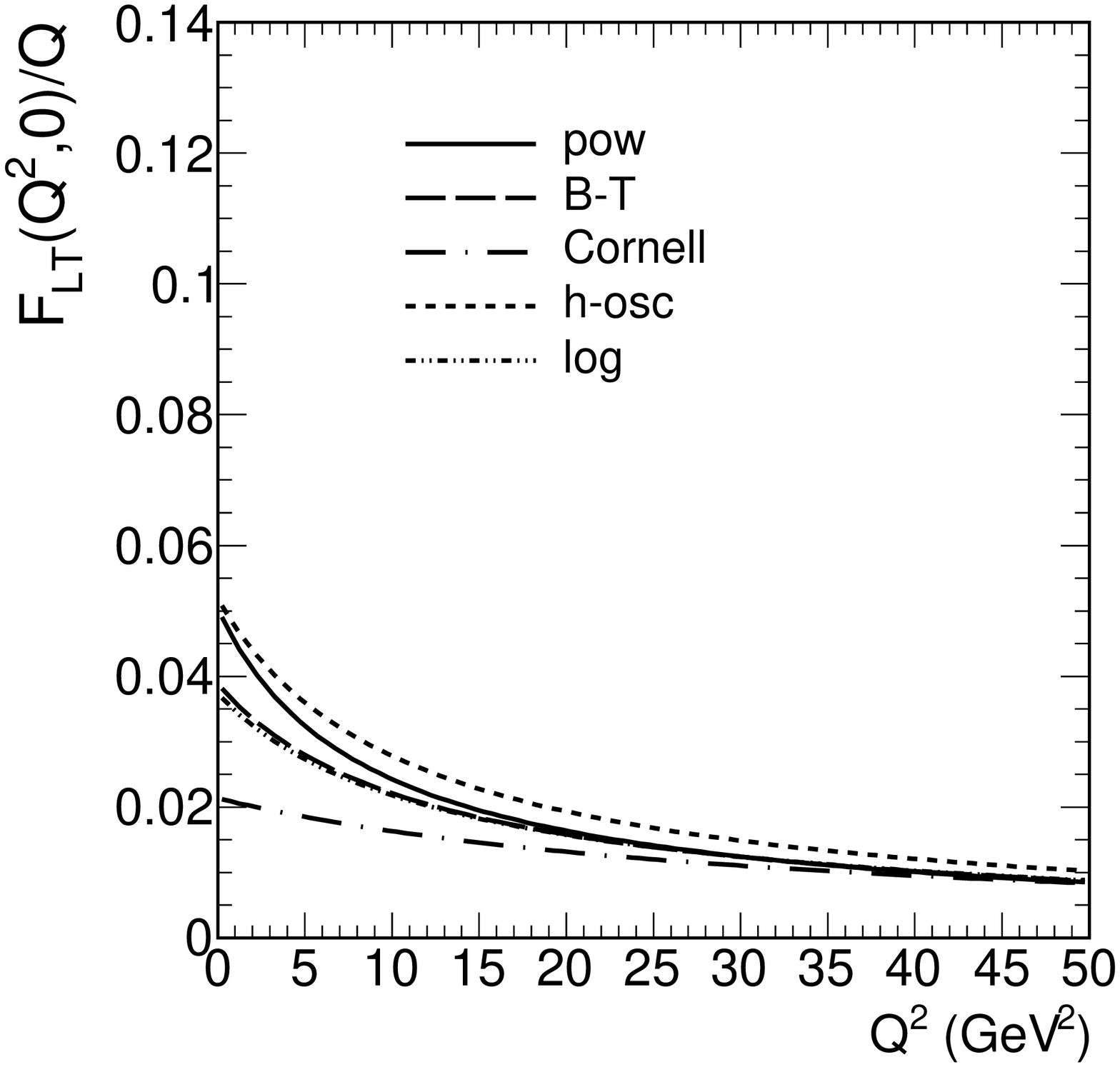}\\
    \includegraphics[width = 0.32\textwidth]{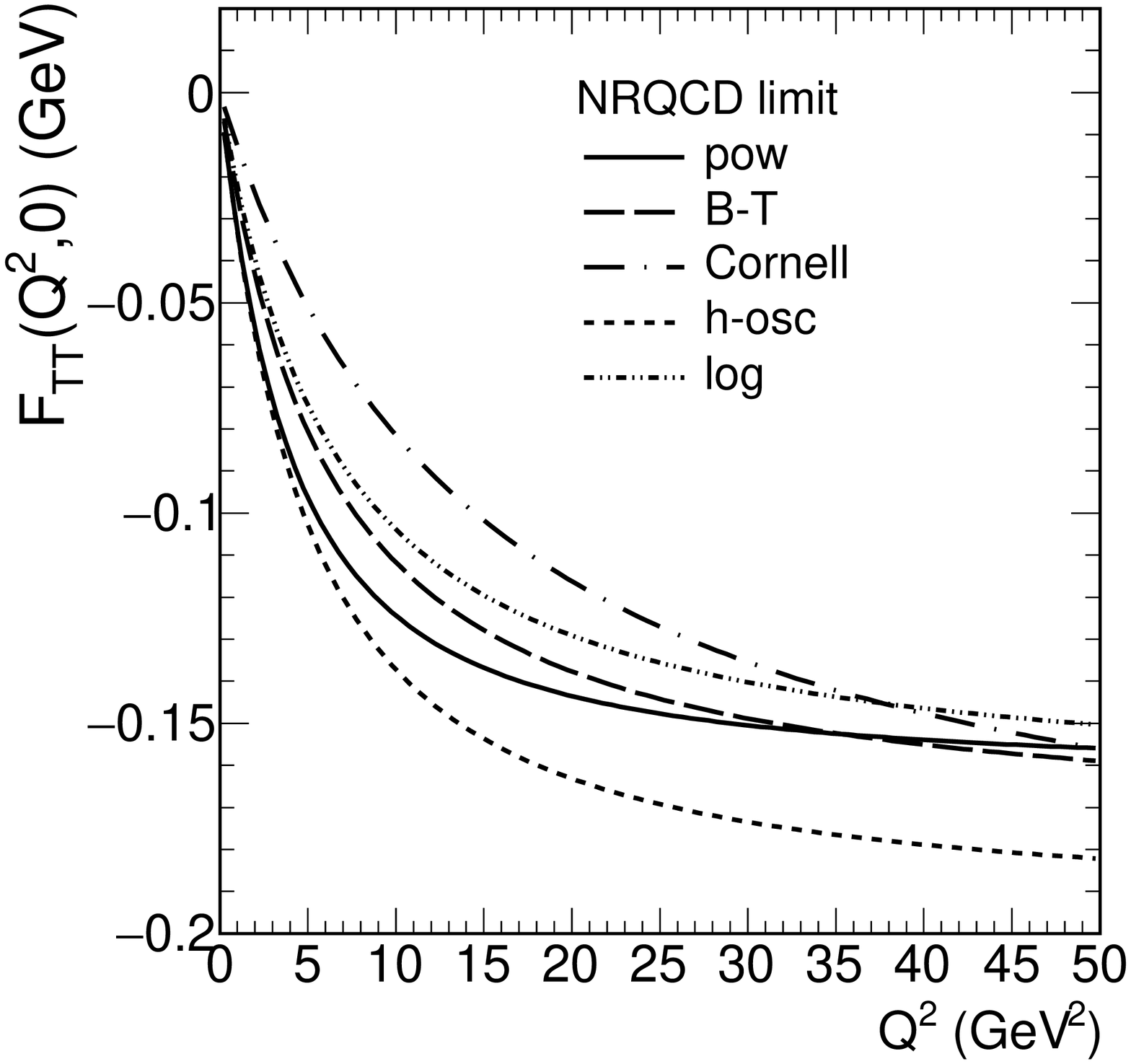}
    \includegraphics[width=0.32\textwidth]{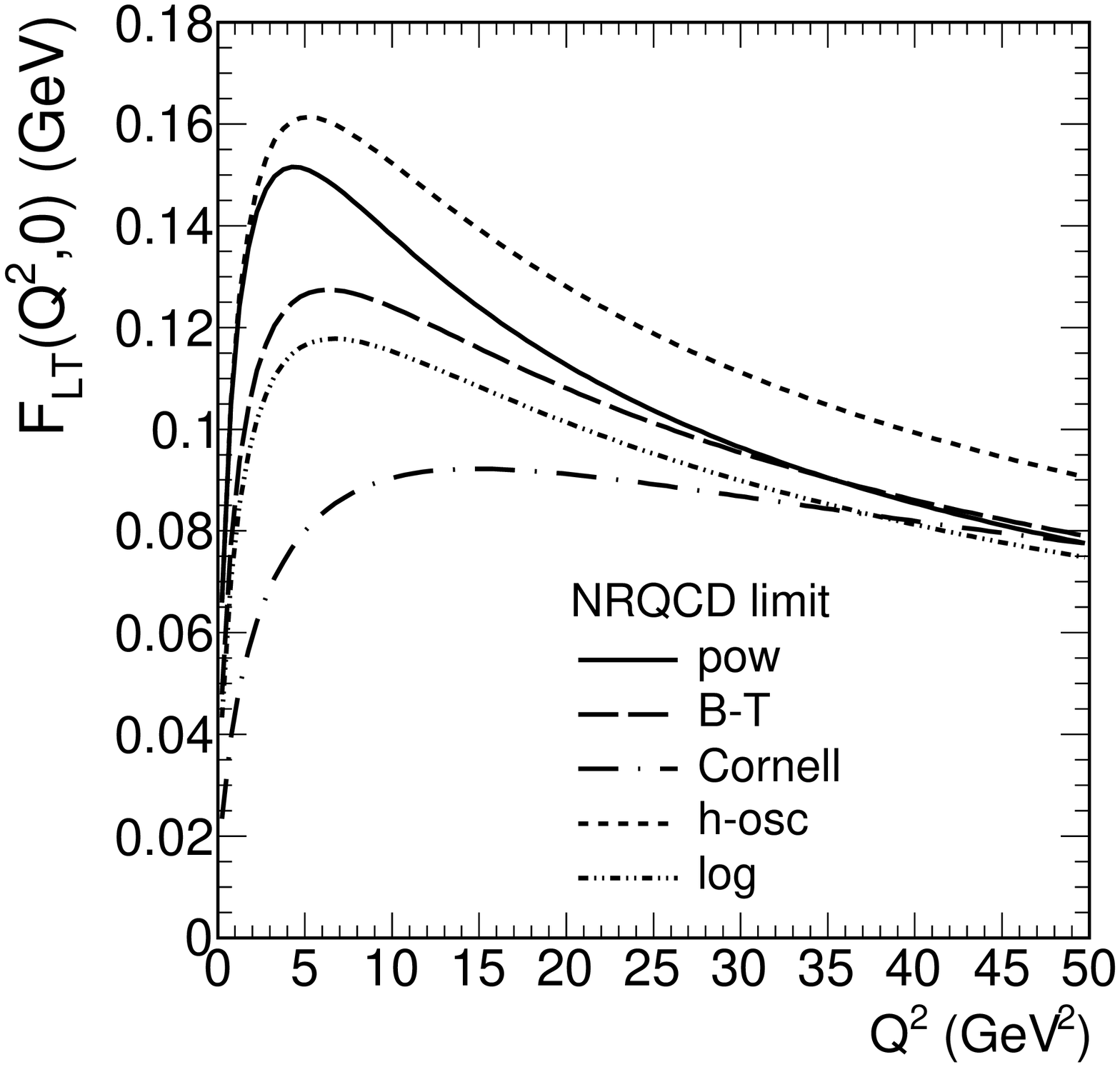}
    \includegraphics[width = 0.32\textwidth]{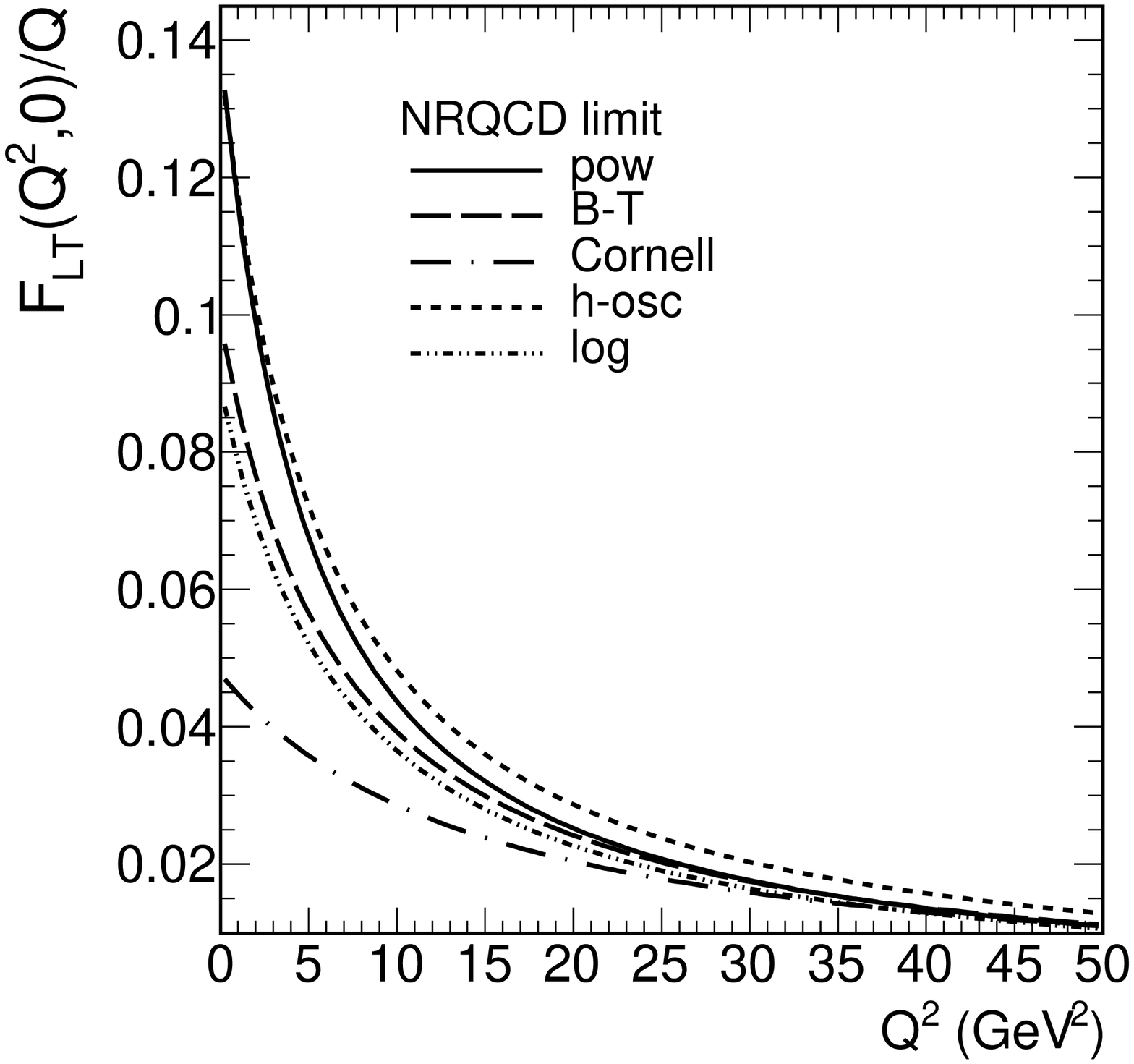}
    \caption{Form factors $F_{\rm TT}(Q^2,0)$, $F_{\rm LT}(Q^2,0)$ for one virtual photon (left and middle panels) and $F_{\rm LT}(Q^2,0)/Q$ (right panel). The top panels represent our results in the LFWF approach and the bottom panels correspond to the nonrelativistic limit.}
    \label{fig:F_TT_F_LT}
\end{figure}

The photon helicity form factors for one virtual and one real photon are shown in Fig.~\ref{fig:F_TT_F_LT}. Here the top row shows the result for the full LCWF, while the bottom row displays the NRQCD limit results. In the rightmost panel, we show $F_{\rm LT}(Q^2,0)/Q$. As we discussed in Sec.~\ref{sec:tensor-dec} it has a finite limit at $Q^2 \to 0$. This value at $Q=0$ is directly related to the so-called reduced width which we will introduce below. In general, we observe a strong difference in the $Q^2$--dependence of form factors between the full calculation and the NRQCD limit. Also different potentials give rise to different shapes of form factors. This makes a future measurement in single-tagged $e^+ e^-$ collisions very interesting. We now turn to the observables of such a measurement.

\subsection{Virtual photon cross sections}

The transition form factors obtained in this work can be used to make predictions for the production of axial-vector mesons in the $e^+ e^- \to e^+ e^- A$ reaction which proceeds via the $\gamma \gamma$ mode. Here one distinguishes between double-tagged processes, where both final state leptons are registered and single-tagged production, where only one of the leptons is measured. The former reactions allow access to the full dependence on $Q_1^2, Q_2^2$ of form factors, while for the latter ones 
one of photons will be quasi-real, and only the one related to the tagged lepton will have finite virtuality $Q^2$.

It is a standard procedure to write the cross section for the $e^+ e^-$ reaction in terms of a density matrix of photon fluxes and virtual photon cross sections as well as response functions related to interferences, see e.g.~Ref.~\cite{Budnev:1975poe,Poppe:1986dq}.

The non-vanishing virtual photon cross sections for the production of an axial-vector resonance read 
\begin{eqnarray}
\sigma_{\rm TT} &=& { 1 \over 4\sqrt{X}} \, \Big( W(++,++) + W(+-,+-) \Big) = {1 \over 4 \sqrt{X}} \, W(++,++) \, ,  \, \, \nonumber \\\
\sigma_{\rm LT}  &=& {1 \over 2 \sqrt{X}} W(0+,0+) \, , \qquad 
\sigma_{\rm TL} = {1 \over 2 \sqrt{X}} W(+0,+0) \,  ,  
\label{eq:sigma_gamma_gamma}
\end{eqnarray}
where for the resonance of mass $M$ and width $\Gamma$ we can write 
\begin{eqnarray}
W(\lambda_1 \lambda_2, \lambda_3 \lambda_4) = {M\Gamma \over (\hat s - M^2)^2 + M^2 \Gamma^2} \sum_{\lambda_A}  {\cal M}^*_{\lambda_1 \lambda_2}    {\cal M}_{\lambda_3 \lambda_4} \, . 
\end{eqnarray}
Here $\sqrt{\hat s}$ is the $\gamma^* \gamma^*$ cm energy, and ${\cal M}_{\lambda_1\lambda_2}$ are the helicity amplitudes in the $\gamma^* \gamma^*$ cm frame. We have collected the latter in Appendix \ref{sec:helicity_cms}. For the narrow resonance as the $\chi_{c1}$ of this work, the Breit-Wigner prefactor above can be replaced by $\pi \delta(\hat s - M^2)$.

Current experiments are not able to perform the $\rm TT$ vs $\rm LT,TL$ separation, and when tagging, say, the first lepton, one is sensitive to the total $\gamma^* \gamma$ cross section
\begin{eqnarray}
\sigma_{\rm tot}(Q^2,0) \equiv \sigma_{\rm TT}(Q^2,0) + \epsilon \sigma_{\rm LT}(Q^2,0)  = \epsilon \sigma_{\rm LT}(Q^2,0) \Big( 1 + {1 \over\epsilon} R_{\rm TL}(Q^2) \Big) \, .
\label{eq:sigma_tot}
\end{eqnarray}
with the photon polarization parameter $\epsilon$, which in practice is close to one, $\epsilon \approx 1$. Now notice, that we have shown that for the $Q \bar Q$ bound state $F_{\rm TT}$ is calculable from $F_{\rm LT}$ and $F_{\rm TL}$. In particular, for the case of one real photon, we have
\begin{eqnarray}
F_{\rm TT}(Q^2,0) = -{Q \over M} F_{\rm LT}(Q^2,0) \, ,
\end{eqnarray}
so that we obtain the universal result
\begin{eqnarray}
R_{\rm TL} = {  F^2_{\rm TT}(Q^2,0) \over 2 F^2_{\rm LT}(Q^2,0)} = {Q^2 \over 2 M^2} \, , 
\label{eq:R_TL}
\end{eqnarray}
which does not depend on the wave function parameters/potential model used.

While the $1^{++}$ meson does not decay into two real photons, it is common practice to introduce (convention-dependent) effective virtual photon widths, see e.g.~Refs.~\cite{TPCTwoGamma:1988izb,Olsson:1987jk}. These are defined through their relation to virtual photon cross sections \emph{defined} through Eq.~( \ref{eq:sigma_gamma_gamma}). We follow the convention\footnote{We note that different experiments historically have used definitions of $ \Gamma^{\rm LT}$ differing by a factor of two \cite{Olsson:1987jk}.} of Ref.~\cite{TPCTwoGamma:1988izb}, where for a resonance of spin $J$, mass $M$ and width $\Gamma$:
\begin{eqnarray}
\sigma_{ij} = {32 \pi (2J+1) \over N_i N_j} {\hat s \over 2M \sqrt{X}} \, {M\Gamma \over (\hat s - M^2)^2 + M^2 \Gamma^2} \, \Gamma^{ij}_{\gamma^* \gamma^*} (Q_1^2,Q_2^2, \hat s)   \, ,
\end{eqnarray}
where $\{ i,j \} \in \{ \rm {T,L} \}$, and $N_{\rm T}=2, N_{\rm L} =1$ are the numbers of polarization states of photons. In terms of our helicity form factor, we obtain for the $\rm LT$ configuration, putting at the resonance pole $\hat s \to M^2$, and $J=1$ for the axial-vector meson:
\begin{eqnarray}
\Gamma^{\rm LT}_{\gamma^* \gamma^*} (Q_1^2,Q_2^2,M^2) = 
{\pi \alpha_{\rm em}^2 \over 3 M} \,  \, F_{\rm LT}^2(Q_1^2,Q_2^2) \, . 
\end{eqnarray}
Of particular interest is the so-called reduced width defined as\footnote{Note, that if one wants to define the reduced width by calculating a $A \to \gamma^*_{\rm L} \gamma$ ``decay'', one ought to introduce an additional factor of $1/2$ in order to agree with the convention of Ref.~\cite{TPCTwoGamma:1988izb}, see Ref.~\cite{Pascalutsa:2012pr,Hoferichter:2020lap}.}
\begin{eqnarray}
\tilde \Gamma (A) = \lim_{Q^2 \to 0} {M^2 \over Q^2} \Gamma^{\rm LT}_{\gamma^* \gamma^*} (Q^2,0,M^2)  = { \pi \alpha^2_{\rm em } M\over 3 } \, f^2_{\rm LT}\, , \, \, \,  {\rm with} \, \, f_{\rm LT} = \lim_{Q^2 \to 0} {F_{\rm LT}(Q^2,0) \over Q} \, , 
\label{eq:reduced_width}
\end{eqnarray}
which provides a useful measure of size of the relevant $e^+ e^-$ cross section in the $\gamma \gamma$ mode.

The coupling strength $f_{\rm LT}$ can be calculated from the LFWF as
\begin{eqnarray}
f_{\rm LT} = - e_f^2 4 M^2 \sqrt{3 N_c \over 2} \int {dz d^2 \bk \over 16 \pi^3} \psi(z,\bk) {\bk^2 \over [\bk^2 + m_Q^2]^2} \, .
\end{eqnarray}
Using the relation of the radial LFWF to the radial WF $u(k)$, one can reduce this further to a one-dimensional integral
\begin{eqnarray}
f_{\rm LT} = - e_f^2 M^2 {\sqrt{3 N_c} \over 8 \pi}  \int_0^\infty {dk \, k^2 u(k) \over (k^2 + m_Q^2)^2} {1 \over \sqrt{M_{Q\bar Q}}} \, \Big\{ {2 \over \beta^2} - {1 - \beta^2 \over \beta^3} \, \log\Big( {1+\beta \over 1 - \beta} \Big) \Big\} \, ,
\end{eqnarray}
with 
\begin{eqnarray}
\beta = {k \over \sqrt{k^2 + m_Q^2}} \,, \qquad M_{Q\bar Q} = 2 \sqrt{k^2 + m_Q^2} \, . 
\end{eqnarray}
In the NRQCD limit of $k \ll m_Q$, and hence $\beta \ll 1$, we obtain
\begin{eqnarray}
{f_{\rm LT}|}_{\rm NRQCD} = - e_f^2 {\sqrt{3 N_c M^3} \over 8 \pi} \, {4 \over 3} {1 \over m_Q^4} \int_0^\infty dk \, k^2 u(k) = - e_f^2 \sqrt{3 N_c \over \pi m_Q^5} R'(0) \, , 
\end{eqnarray}
so that the reduced width expressed in terms of the derivative of the WF at the origin reads:
\begin{eqnarray}
\tilde \Gamma(A) = {2 \alpha^2_{\rm em} e_f^4 N_c \over m_Q^4} \, |R'(0)|^2  \, ,
\end{eqnarray}
which is in agreement with the result of Ref.~\cite{Danilkin:2017utg}. Notice, that in the NRQCD limit we have substituted $M = 2m_Q$ everywhere.

The reduced widths for different potentials used are listed in Table~\ref{tab:reduced_width}. We show result in our relativistic treatment as well as in the commonly used NRQCD approximation. These two values differ considerably and strongly depend on the potential used. In general, the full result is smaller than the NRQCD limit. For all potentials $\tilde \Gamma(\chi_{c1})$ is substantially smaller than 1 keV. Considerably larger values of $\tilde \Gamma(\chi_{c1})$ are quoted in the literature. For example, in Ref.~\cite{Danilkin:2017utg} a value of $\tilde \Gamma(\chi_{c1}) \approx 1.6 \, \rm{keV}$ is reported from a sum rule analysis. In Ref.~\cite{Li:2021ejv} a value of $\tilde \Gamma(\chi_{c1}) \approx 3 \, \rm{keV}$ is obtained from a LFWF approach\footnote{Comparing with the helicity amplitudes found in Ref.~\cite{Hoferichter:2020lap} it appears however that Ref.~\cite{Li:2021ejv} uses a definition of the reduced width which is a factor of two larger than ours.}. A measurement of the reduced width would therefore be very valuable.

\begin{figure}
    \centering
    \includegraphics[width = 0.45\textwidth]{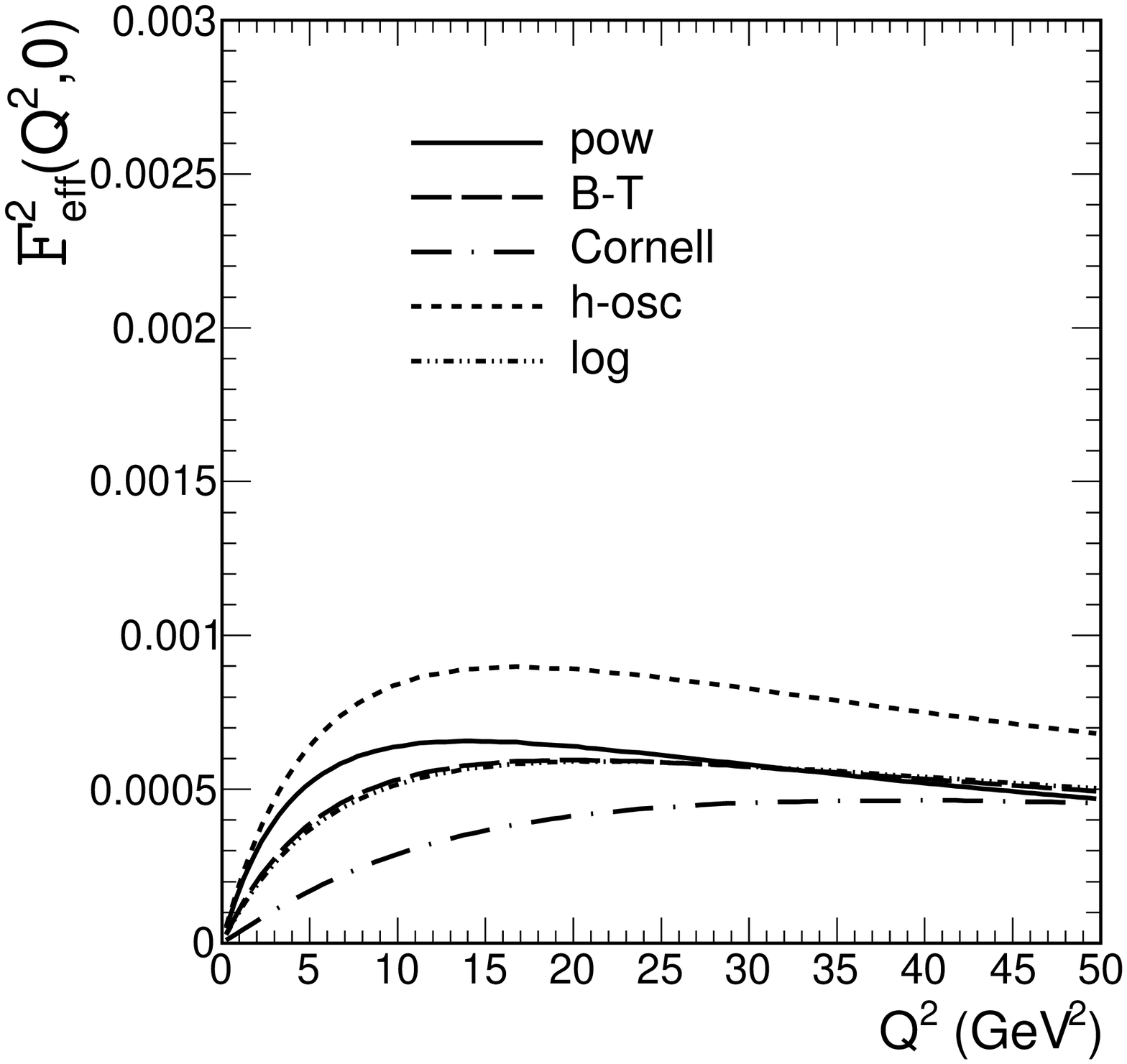}
    \includegraphics[width = 0.45\textwidth]{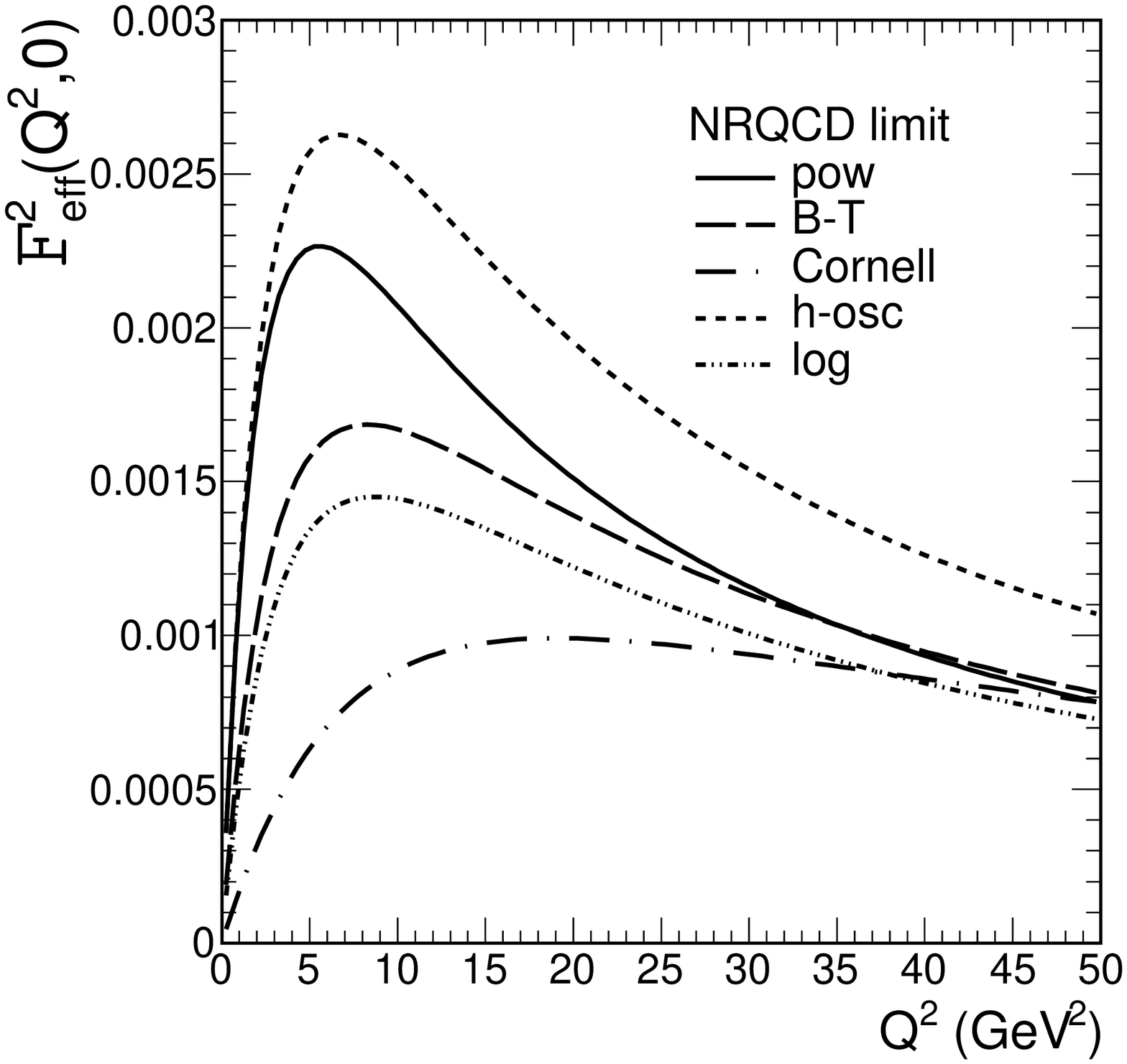}
    \caption{The square of the effective form factor Eq.~(\ref{eq:F_eff}) as a function of photon virtuality within LFWF approach (on the l.h.s.) and in the nonrelativistic limit (on the r.h.s.).}
    \label{fig:FF_eff}
\end{figure}

A comment on the total cross section (see Eq.~(\ref{eq:sigma_tot})) is in order. Using the result of Eq.~(\ref{eq:R_TL}), and putting $\epsilon=1$, one obtains
\begin{eqnarray}
\sigma_{\rm tot}(Q^2,0) &=& 16 \pi^3 \alpha_{\rm em}^2 \delta(\hat s - M^2) \, {Q^2 \over Q^2 + M^2} \Big( 1 + {Q^2 \over 2 M^2} \Big) \, \Big({F_{\rm LT}(Q^2,0) \over Q} \Big)^2 \, \nonumber \\
&\equiv&  16 \pi^3 \alpha_{\rm em}^2 \delta(\hat s - M^2) \, {\texttt{F}}^2_{\rm eff}(Q^2) \, . 
\end{eqnarray}
Here, the dependence on $Q^2$ is controlled by the effective form factor\footnote{See Ref.~\cite{Schuler:1997yw} for a discussion in the NR quark model.} ${\texttt{F}}_{\rm eff}(Q^2)$:
\begin{eqnarray}
{\texttt{F}}_{\rm eff}(Q^2) = \sqrt{Q^2 \over Q^2 + M^2} \, \sqrt{ 1 + {Q^2 \over 2 M^2}} \, {F_{\rm LT}(Q^2,0) \over Q} \, .
\label{eq:F_eff}
\end{eqnarray}
We show plots of the effective form factor squared in Fig.~\ref{fig:FF_eff}. In the left panel we show the results of the full LFWF approach for different potentials, while in the right panel we show the NRQCD results. The two approaches lead to different shapes of the effective formfactor. Also, in the NRQCD limit the formfactor is substantially larger.
\begin{table}[h]
    \caption{Reduced width, see Eq.~(\ref{eq:reduced_width})}
    \begin{tabular}{l|c|c|c|c}
    potential model & $m_c$ (GeV) & $|R'(0)|$ (GeV$^{5/2}$) & $\tilde \Gamma(\chi_{c1})_{\rm NRQCD}$ (keV) & $\tilde \Gamma(\chi_{c1})$ (keV) \\
    \hline
    \hline
    power-law           & 1.33 & 0.22 & 0.97 & 0.50 \\
    Buchm\"uller-Tye    & 1.48 & 0.25 & 0.82 & 0.30 \\
    Cornell             & 1.84 & 0.32 & 0.56 & 0.09 \\
    harmonic oscillator & 1.4  & 0.27 & 1.20 & 0.53 \\
    logarithmic         & 1.5  & 0.24 & 0.72 & 0.27 
    \end{tabular}
    \label{tab:reduced_width}
\end{table}

\section{Conclusions}
\label{sec:Conclusions}

We have studied the $\gamma^* \gamma^* \to 1^{++}$ quarkonium transition form factors in a light front approach. The quarkonium LFWFs have been obtained from solutions of the Schr\"odinger equation for $c \bar c$ interaction transformed to the LF via the Terent'ev prescription for the radial part and a Melosh transform for the spin-orbit part. Different interaction potentials from the literature have been used to generate the charmonium wave function.

In our derivation we started from the $\gamma^* \gamma^*$-fusion amplitude for the light-like photon polarizations $n^+_\mu,n^-_\mu$. Subsequently we have demonstrated how to extract invariant form factors $F_{\rm TT}$, $F_{\rm LT}$, $F_{\rm TL}$, which refer to photon polarizations in the $\gamma^* \gamma^*$ c.m.-frame. We have obtained these form factors as a function of both photon virtualities. We have found that for the $Q \bar Q$--bound state the form factor $F_{\rm TT}$ is determined through $F_{\rm LT}$ and $F_{\rm TL}$. 

We have also analyzed the situation where only one photon is virtual and the second photon is assumed to be real, which is relevant for single-tagged $e^+ e^-$ collisions. Here, the ratio $F_{\rm TT}/F_{\rm LT}$ turned out to be universal and independent of the potential used.

Our predictions for $\gamma^* \gamma^* \to \chi_{c1}$ are ready for experimental verification. This could be done for example by the Belle-2 collaboration in $e^+ e^- \to e^+ e^- \chi_{c1}$ reaction, by measuring e.g. $\chi_{c1} \to J/\psi \gamma$ with one and two electron tagging (${\rm Br}(\chi_{c1} \to J/\psi \gamma)$ = $(34.3\,\pm\,1.0)$~\% \cite{Workman:2022ynf}). A similar study has already been initiated in Ref.~\cite{Belle:2020ndp}  for the axial-vector meson $\chi_{c1}(3872)$ (or X(3872)), using the $X(3872) \to J/\psi \pi^+ \pi^-$ decay. This meson is supposed to have a large non-$c \bar c$ component, and is still  rather enigmatic (see e.g.~Ref.~\cite{Cisek:2022uqx} for a consideration of its production in the gluon-gluon fusion process including a $c \bar c$ admixture).

\vspace{1cm}

{\bf Acknowledgements}

This work was partially supported by the Polish National Science Center grant UMO-2018/31/B/ST2/03537 and by the Center for Innovation and Transfer of Natural Sciences and Engineering Knowledge in Rzesz{\'o}w.
R.P. is partially supported by the Swedish Research Council grant No. 2016-05996, by the European Research Council (ERC) under the European Union's Horizon 2020 research and innovation programme (grant agreement No 668679), as well as by the NKFI grant K133046 (Hungary).

\appendix
\section{Axial-vector meson polarisation states, wave functions}
\label{sec:states}

Here, we would like to summarise the basic formulas for axial-vector meson wave functions for different meson polarisations.

The polarization vectors $\vec E(\lambda_A)$ introduced in Eq.~(\ref{eq:RF_WF}) read
\begin{eqnarray}
\vec E (\pm) = (\bE(\pm),0) \, , \qquad \vec E(0) = \vec n \, , \qquad   
\bE (\lambda_A) = -{1 \over \sqrt{2}} \Big( \lambda_A \be_x + i \be_y \Big) \, .
\end{eqnarray}
The axial-vector meson WF in the $Q\bar Q$ rest frame as given in Eq.~(\ref{eq:RF_WF}) and its radial part $u(k)$ are normalized as
\begin{eqnarray}
\int d^3 \vec k \,  \sum_{\tau \bar \tau}  \Psi^{(\lambda'_A)*}_{\tau \bar \tau} (\vec k) \Psi^{(\lambda_A)}_{\tau \bar \tau} (\vec k) = \delta_{\lambda'_A \lambda_A}  \, , \qquad \int_0^\infty dk \, u^2(k) = 1 \, .
\end{eqnarray}
The normalization condition for the LFWF reads accordingly
\begin{eqnarray}
\int {dz d^2 \bk \over z (1-z) 16 \pi^3} \,  \sum_{\lambda \bar \lambda}  \Psi^{(\lambda'_A)*}_{\lambda \bar \lambda} (z,\bk) \Psi^{(\lambda_A)}_{\lambda \bar \lambda} (z,\bk) = \delta_{\lambda'_A \lambda_A} \, , 
\end{eqnarray}
while the ``radial WF'' is normalized as
\begin{eqnarray}
\int {dz d^2\bk \over z (1-z) 16 \pi^3} 2 (M_{Q \bar Q}^2 - 4 m_Q^2) \, \psi^2(z,\bk) = 1 \, .
\end{eqnarray}
Finally, we recall that the derivative of the radial wave function at the spatial origin is related to the momentum-space WF $u(k)$ via
\begin{eqnarray}
\int_0^\infty dk \, k^2 u(k) = 3 \sqrt{\pi \over 2} R'(0) \, . 
\end{eqnarray}

The Melosh transform in Eq.~(\ref{eq:Melosh}) is performed straightforwardly e.g.~by using Eq.~(A.15) from Ref.~\cite{Babiarz:2020jkh}. 

Below, we collect the explicit representation of the LFWF for different meson polarizations in the following matrix form:
\begin{eqnarray}
\Psi^{(\lambda_A)}_{\lambda \bar \lambda}(z,\bk) =
\begin{pmatrix}
\Psi^{(\lambda_A)}_{++}(z,\bk) & \Psi^{(\lambda_A)}_{+-}(z,\bk) \\
\Psi^{(\lambda_A)}_{-+}(z,\bk) & \Psi^{(\lambda_A)}_{--}(z,\bk) 
\end{pmatrix} \, .
\end{eqnarray}

\begin{subsection}{Transversely polarized states, $ \lambda_A = \pm 1$}

	\begin{eqnarray}
		\Psi^{(\lambda_A)}_{\lambda \bar \lambda}(z,\bk) &=&
		{\psi(z,\bk) \over \sqrt{z(1-z)}} \sqrt{{3\over 2}}
		\scalemath{0.9}{\begin{pmatrix}
		m_Q(1-2z) \sqrt{2} i[\be(-),\bE(\lambda_A)]  & -(1-2z) (\bE(\lambda_A)\bk) + i [\bE(\lambda_A),\bk] \\
		(1-2z) (\bE(\lambda_A)\bk) + i [\bE(\lambda_A),\bk] &   m_Q(1-2z) \sqrt{2} i[\be(+),\bE(\lambda_A)]
		\end{pmatrix}},\nonumber \\
	\end{eqnarray}
Let us add some useful combinations:
	\begin{eqnarray}
		\Psi^{*(\lambda_A)}_{+-}(z,\bk) - \Psi^{*(\lambda_A)}_{-+}(z,\bk)
		&=& \sqrt{{3 \over 2}} { \psi(z,\bk) \over \sqrt{z (1-z)}} \, 2 (2z-1) (\bE^*(\lambda_A) \bk) \,, \nonumber \\
		\Psi^{*(\lambda_A)}_{+-}(z,\bk) + \Psi^{*(\lambda_A)}_{-+}(z,\bk)
		&=& \sqrt{{3 \over 2}} { \psi(z,\bk) \over \sqrt{z (1-z)}}
		\, (-2i) [\bE^*(\lambda_A), \bk] \,, \nonumber \\
		\sqrt{2} (\be(-)\bq_1) \Psi^{*(\lambda_A)}_{++}(z,\bk)
		+\sqrt{2} (\be(+)\bq_1) \Psi^{*(\lambda_A)}_{--}(z,\bk) &=&
		\sqrt{{3 \over 2}} { \psi(z,\bk) \over \sqrt{z (1-z)}}
		\, 2m_Q (2z-1) i [\bq_1,\bE^*(\lambda_A)] \,.
		\nonumber \\
	\end{eqnarray}
\end{subsection}

\begin{subsection}{Longitudinally polarized  states, $\lambda_A = 0$}

	\begin{eqnarray}
		\Psi^{(0)}_{\lambda \bar \lambda}(z,\bk) &=& {\psi(z,\bk) \over \sqrt{z(1-z)}} \sqrt{{3\over 2}} { 1 \over M_{Q \bar Q}} 
		\begin{pmatrix}
		i 2m_Q \sqrt{2} [\be(-), \bk] & - 2 \bk^2 \\
		2 \bk^2 &  i 2m_Q \sqrt{2} [\be(+),\bk]
		\end{pmatrix} \; .
	\end{eqnarray}
We list here the following combinations relevant for calculation of the production amplitude:
	\begin{eqnarray}
		\Psi^{*(\lambda_A)}_{+-}(z,\bk) - \Psi^{*(\lambda_A)}_{-+}(z,\bk)
		&=&  \sqrt{{3 \over 2}} { \psi(z,\bk) \over \sqrt{z (1-z)}}
		{1 \over M_{Q \bar Q}} (-4 \bk^2) \, , \nonumber \\
		\Psi^{*(\lambda_A)}_{+-}(z,\bk) + \Psi^{*(\lambda_A)}_{-+}(z,\bk)
		&=& 0 \,,\nonumber \\
		\sqrt{2} (\be(-)\bq_1) \Psi^{*(\lambda_A)}_{++}(z,\bk)
		+\sqrt{2} (\be(+)\bq_1) \Psi^{*(\lambda_A)}_{--}(z,\bk) &=&
		\sqrt{{3 \over 2}} { \psi(z,\bk) \over \sqrt{z (1-z)}}
		{2m_Q \over M_{Q \bar Q}} \, (-2i) [\bq_1,\bk] \, . \nonumber \\
	\end{eqnarray}
		
	\end{subsection}

\section{Integrals for numerical evaluation}
\label{sec:integrals}

For transverse polarizations of the meson, the integrals appearing in Eqs.~(\ref{Tpm}), (\ref{J-12}) and (\ref{J-34}) are given in the following explicit form:
\begin{eqnarray}
J_0(\bq_1^2,\bq_2^2) &=& -4 \sqrt{3 \over 2} \int {dz d^2\bk \over z (1-z) 16 \pi^3} \psi(z,\bk) m_Q^2 (1-2z) \Big( {1 \over D_A} - {1 \over D_B} \Big) \, , \nonumber \\
J_1(\bq_1^2,\bq_2^2) &=& -8 \sqrt{3 \over 2} \int {dz d^2\bk \over z (1-z) 16 \pi^3} \psi(z,\bk) z(1-z) {\bq_2 \cdot \bk \over \bq_2^2} \Big( {1 \over D_A} - {1 \over D_B} \Big) \, , \nonumber \\
J_2(\bq_1^2,\bq_2^2) &=& -4 \sqrt{3 \over 2} \int {dz d^2\bk \over z (1-z) 16 \pi^3} \psi(z,\bk) (1-2z) {\bq_2 \cdot \bk \over \bq_2^2} \Big( {1-z \over D_A} + {z \over D_B} \Big) \, , \nonumber \\
J_3(\bq_1^2,\bq_2^2) &=& -4 \sqrt{3 \over 2} \int {dz d^2\bk \over z(1-z) 16 \pi^3} \psi(z,\bk) (1-2z)
{\bq_2^2 \bk^2 - (\bq_2 \cdot \bk)^2 \over \bq_2^2}
\Big( {1 \over D_A} - {1 \over D_B} \Big) \, , \nonumber \\
J_4(\bq_1^2,\bq_2^2) &=& -4 \sqrt{3 \over 2} \int {dz d^2\bk \over z(1-z) 16 \pi^3} \psi(z,\bk) (1-2z)
{ (\bq_2 \cdot \bk)^2 \over \bq_2^2 } \Big( {1 \over D_A} - {1 \over D_B} \Big ) \, ,
\end{eqnarray}
and should be evaluated numerically.

\section{Helicity amplitudes in the $\gamma^* \gamma^*$-c.m. frame and virtual photon cross sections}
\label{sec:helicity_cms}

In the $\gamma^* \gamma^*$ cm frame, we take the momenta of
colliding photons along the $z$-axis:
\begin{eqnarray}
q_{1\mu} = \omega_1 \hat t_\mu + q_z \hat z_\mu \, , \qquad
q_{2\mu} = \omega_2 \hat t_\mu - q_z \hat z_\mu \, , 
\end{eqnarray}
with 
\begin{eqnarray}
\omega_1 = {M^2 + Q_2^2 - Q_1^2 \over 2M} \, , \qquad \omega_2 = {M^2 + Q_1^2 - Q_2^2 \over 2 M}\, , \qquad q_z = {\sqrt{X} \over M} \, ,
\end{eqnarray}
and for the covariantly defined time and $z$-directions:
\begin{eqnarray}
\hat t_\mu = {q_{1\mu} + q_{2\mu} \over M} \, , \qquad \hat z_\mu = {M \over 2 \sqrt{X}} \Big( q_{1\mu} - q_{2\mu} + {Q_1^2 - Q_2^2 \over M^2} ( q_{1\mu} + q_{2\mu} ) \Big) \, .
\end{eqnarray}
Let $\hat x_{\mu}, \hat y_{\mu}$ be two spacelike unit vectors $\hat x^2 = \hat y^2 = -1$ orthogonal to $\hat z_\mu$. Then we can write
\begin{eqnarray}
\tilde G_{\mu \nu} = -\sqrt{X} \epsilon_{\mu \nu \rho \sigma} \hat t_\rho \hat z_\sigma  = - \sqrt{X} \Big( \hat x_\mu \hat y_\nu - \hat x_\nu \hat y_\mu \Big) \, .
\end{eqnarray}
Therefore, our amplitude of Eq.~(\ref{eq:formfactors}) takes the intuitively obvious form
\begin{eqnarray}
{1 \over 4 \pi \alpha_{\rm em}}{\cal M}_{\mu \nu \rho} &=&  \, 
\, - i \Big( \hat x_\mu \hat y_\nu - \hat x_\nu \hat y_\mu \Big) \hat z_\rho 
F_{\rm TT}(Q_1^2,Q_2^2) \nonumber\\
&-& i  e_\mu^L(q_1) \,   \Big( \hat x_\nu \hat y_\rho - \hat x_\rho \hat y_\nu \Big)
 F_{\rm LT}(Q_1^2,Q_2^2) 
- i e_\nu^L(q_2) 
 \Big( \hat x_\mu \hat y_\rho - \hat x_\rho \hat y_\mu \Big) F_{\rm TL}(Q_1^2,Q_2^2) \, . \nonumber \\
\label{eq:formfactors_RF}
\end{eqnarray}
For completeness, we introduce the polarization vectors of the axial-vector meson $E_\mu(\lambda_A)$ and photons $e_\mu^{1,2}(\lambda_{1,2})$ in the $\gamma \gamma$-frame as
\begin{eqnarray}
E_\mu(0) = \hat z_\mu \, , \qquad E_\mu(\pm) = e^1_\mu(\pm) =- {1 \over \sqrt{2}} ( \pm \hat x_\mu + i \hat y_\mu ) \, , \qquad e^2_\mu(\pm) = e^1_\mu(\mp) \, ,  
\end{eqnarray}
and for longitudinal photons
\begin{eqnarray}
e^1_\mu(0) = e^L_\mu (q_1)  = {1 \over Q_1} ( q_z \hat t_\mu + \omega_1 \hat z_\mu) \,, \qquad 
e^2_\mu(0) = e^L_\mu(q_2) = {1 \over Q_2} ( - q_z \hat t_\mu + \omega_2 \hat z_\mu) \, .
\end{eqnarray}
Then, the helicity amplitudes in the $\gamma \gamma$-frame are obtained from
\begin{eqnarray}
{\cal M}_{\lambda_1 \lambda_2} = e^{1\mu}(\lambda_1) e^{2\nu}(\lambda_2) \, {\cal M}_{\mu \nu \rho} E^{\rho *}(\lambda_A) \, , \qquad \lambda_A = \lambda_1 - \lambda_2 \, . 
\end{eqnarray}
We can now easily read off the helicity amplitudes as
\begin{eqnarray}
{\cal M}_{++} =  4 \pi \alpha_{\rm em}  F_{\rm TT}(Q_1^2, Q_2^2) \, , \qquad
{\cal M}_{0+} =  - 4 \pi \alpha_{\rm em} F_{\rm LT}(Q_1^2,Q_2^2) \, . 
\end{eqnarray}
This is the easiest way to relate our form factors to other conventions in the literature.

Finally, let us for convenience quote the results in the NRQCD limit:
\begin{eqnarray}
{\cal M}_{++} &=& 8 \,  \alpha_{\rm em} e_f^2  \sqrt{6 N_c \pi \over M^3} \, R'(0) {Q_1^2 - Q_2^2 \over \nu} \, , \qquad
{\cal M}_{0+} = 8 \alpha_{\rm em} e_f^2 \sqrt{6 N_c \pi \over M} \, R'(0) \,  {Q_1(\nu + Q_2^2 )\over \nu^2} \, . \nonumber \\  
\end{eqnarray}
These results are in agreement with \cite{Schuler:1997yw,Danilkin:2017utg}.

\section{Projection on linear polarizations}
\label{sec:linear_pol} 

We define the covariant tetrad associated with the LF polarizations as 
\begin{eqnarray}
\hat t_\mu = {P_\mu \over M} \, , \quad \hat z_\mu = E_\mu(0) \, , \quad
\hat x_\mu = {P^\perp_\mu \over |\bP|} + {|\bP| \over P_+} n^-_\mu \, , \quad \hat y_\mu = - \varepsilon_{\mu \nu \rho \sigma} \hat x^\nu \hat t^\rho \hat z^\sigma \, .
\end{eqnarray}
The projection of the amplitude on $z$-direction then reads:
\begin{eqnarray}
{\cal M}(0) \equiv {\cal M}_z &=&  {-i M \over X} [\bq_1,\bq_2] \Big\{ {M^2 + Q_1^2 - Q_2^2 \over 2 M} F_{\rm TT}(Q_1^2, Q_2^2)  + Q_2 F_{\rm TL}(Q_1^2, Q_2^2) \Big \}  \nonumber \\
&=& {-i  \over 2X} [\bq_1,\bq_2]
\Big\{ (M^2 + Q_1^2 - Q_2^2) F_{\rm TT}(Q_1^2, Q_2^2) + F_A(Q_1^2, Q_2^2) + F_S(Q_1^2,Q_2^2) \Big\} \nonumber \\
&=& {-i  \over 2X} Q_1Q_2 \sin \phi
\Big\{ (Q_1^2 - Q_2^2) F_{\rm TT}(Q_1^2,Q_2^2) + M F_S(Q_1^2,Q_2^2) \nonumber \\
&+& M^2 F_{\rm TT}(Q_1^2,Q_2^2) + M F_A(Q_1^2,Q_2^2) \Big \} \, .
\end{eqnarray}
We notice that it has both a symmetric and antisymmetric piece under $Q_1^2 \leftrightarrow Q_2^2$ interchange.

Now we turn to the projection on the $x$-axis:
\begin{eqnarray}
{\cal M}_x &=& {i \over X |\bP|} [\bq_1,\bq_2]
\Big \{ Q_2^2 M F_{\rm TT}(Q_1^2,Q_2^2) 
+(Q_1^2 + Q_2^2) {1 \over 2} F_A(Q_1^2,Q_2^2) -{1 \over 2} M^2 F_S(Q_1^2,Q_2^2) \nonumber \\
&+& (\bq_1 \cdot \bq_2) \Big( M F_{\rm TT}(Q_1^2,Q_2^2) + F_A(Q_1^2,Q_2^2) \Big) \Big \} \nonumber \\
&=& i {Q_1 Q_2 \over 2 X |\bP|}
\Big\{ \sin \phi \Big(
2 Q_2^2 M F_{\rm TT}(Q_1^2,Q_2^2) 
+(Q_1^2 + Q_2^2) F_A(Q_1^2,Q_2^2) - M^2 F_S(Q_1^2,Q_2^2)\Big) \nonumber \\
&+& Q_1 Q_2 \sin (2 \phi) \Big( M F_{\rm TT}(Q_1^2, Q_2^2) + F_A(Q_1^2,Q_2^2) \Big) \Big \} \nonumber  \\
&=&  i {Q_1 Q_2 \over 2 X |\bP|}
\Big\{ \sin \phi \Big[ - M (Q_1^2 - Q_2^2) F_{\rm TT}(Q_1^2,Q_2^2) -  
M^2 F_S(Q_1^2,Q_2^2) \nonumber \\
&+& (Q_1^2 + Q_2^2) \Big( M  F_{\rm TT}(Q_1^2,Q_2^2) + F_A(Q_1^2,Q_2^2) \Big) \Big]  \nonumber \\
&+&  Q_1 Q_2 \sin (2 \phi) \Big( M F_{\rm TT}(Q_1^2, Q_2^2) + F_A(Q_1^2,Q_2^2) \Big) \Big \} \, .
\end{eqnarray}

Finally, the projection onto the $y$-axis (normal to the production plane) is given by
\begin{eqnarray}
{\cal M}_y &=& {i \over 2 X |\bP|}
\Big\{ Q_1^2 Q_2^2 \Big( F_A(Q_1^2,Q_2^2) - M F_{\rm TT}(Q_1^2,Q_2^2) \Big) \nonumber \\
&+& \half (M^2 + Q_1^2 + Q_2^2) \Big[ (Q_1^2 - Q_2^2) F_S(Q_1^2, Q_2^2) + 
(Q_1^2 + Q_2^2) F_A(Q_1^2,Q_2^2) \Big] \nonumber \\
&+&  Q_1 Q_2 \cos \phi \Big[
(M^2 + 2 Q_1^2 + 2 Q_2^2) F_A(Q_1^2,Q_2^2) + (Q_1^2 - Q_2^2) F_S(Q_1^2,Q_2^2)  \Big] \nonumber \\
&+&  Q^2_1 Q^2_2 \cos(2 \phi) 
\Big[ F_A(Q_1^2,Q_2^2) + M F_{\rm TT}(Q_1^2,Q_2^2) 
\Big] \Big \} \, .
\end{eqnarray}

The $x,z$ components contain both symmetric and antisymmetric parts. Indeed, these components contain only two combinations of form factors:
\begin{eqnarray}
f_S(Q_1^2,Q_2^2) &=& (Q_1^2 - Q_2^2) 
F_{\rm TT}(Q_1^2,Q_2^2) + M F_S(Q_1^2,Q_2^2)\,, \nonumber \\
f_A(Q_1^2,Q_2^2) &=& M F_{\rm TT}(Q_1^2,Q_2^2) + F_A(Q_1^2,Q_2^2) \, , 
\end{eqnarray}
which are symmetric and antisymmetric w.r.t. the interchange $Q_1^2 \leftrightarrow Q_2^2$, respectively. The $x$ and $z$ components therefore are not enough to fix the three functions $F_{\rm TT},F_A,F_S$. 

This is different, however, for the $y$-component. We note that the $y$-component is fully antisymmetric w.r.t. interchange $Q_1^2 \leftrightarrow Q_2^2$. Here, we have three orders in the Fourier expansion in $\cos n \phi$, which would give us three equations for the three form factors.

\bibliography{bib}

\end{document}